\documentclass[notitlepage]{scrartcl}

\setkomafont{caption}{\footnotesize}
\setkomafont{captionlabel}{\bfseries}

\usepackage{packages}

\usepackage{commands}

\title{Dimension-Reduced Cosine Expansions via Tensor Decomposition: Methodology and Applications to Credit Exposure Quantification}

\author{
    Gijs Mast\thanks{Delft Institute of Applied Mathematics, Delft University of Technology, 2628 CD Delft, the Netherlands
(\href{mailto:G.L.Mast@tudelft.nl}{G.L.Mast@tudelft.nl}). Part of this research was conducted during this author’s affiliation with \href{https://fsquaredquant.nl}{FF Quant Advisory B.V.}} 
    \and
    Fang Fang\thanks{\href{https://fsquaredquant.nl}{FF Quant Advisory B.V.}, 3531 WR Utrecht, the Netherlands (\href{mailto:fang.fang@ffquant.nl}{fang.fang@ffquant.nl}) and Delft Institute of Applied Mathematics, Delft University of Technology, 2628 CD Delft, the Netherlands  ( \href{mailto:f.fang@tudelft.nl}{f.fang@tudelft.nl}).}
    \and
    Xiaoyu Shen\thanks{\href{https://fsquaredquant.nl}{FF Quant Advisory B.V.}, 3531 WR Utrecht, the Netherlands (\href{mailto:xiaoyu.shen@ffquant.nl}{xiaoyu.shen@ffquant.nl}).} 
    \and
    Marnix Brands\thanks{\href{https://fsquaredquant.nl}{FF Quant Advisory B.V.}, 3531 WR Utrecht, the Netherlands.} 
}

\begin{document}
\maketitle
\begin{abstract}
Monte Carlo (MC) simulation is the standard method in the banking industry for computing credit exposures due to its flexibility and general applicability. For large portfolios, however, the estimation of tail risks— such as high quantiles of Potential Future Exposure (PFE) — is challenging, as the associated estimators may exhibit large variance. Accurate resolution of extreme scenarios therefore requires a substantial number of simulation paths, which is computationally prohibitive. In practice, computational budgets often limit the number of paths, which can impair the stability of tail estimates and has motivated increased supervisory scrutiny of such models. 

This paper initiates a series of studies on a \emph{COS--tensor} framework, as an efficient alternative to MC for large and liquid portfolios characterized by a modest number of dominant risk factors but a large number of trades.
The framework is built on three key insights. First, the cumulative distribution function (CDF) of portfolio level mark-to-market (MtM) values and exposures can be recovered in the Fourier domain by first numerically evaluating their characteristic functions and subsequently applying the
one-dimensional COS method~\cite{fang2009novel}. Second, the curse of dimensionality arising in the evaluation of netting-set characteristic functions is mitigated by \emph{dimension-reduced cosine expansions} derived  by combining Fourier--cosine series representations with tensor decomposition techniques. This reformulation shifts the main computational burden from online evaluation to an offline training stage. Third, the offline training can be performed directly in the Fourier domain by reapplying the core idea of the COS method. This improves both training speed and accuracy by more than two orders of magnitude compared with gradient‑based training in the physical domain.

This paper establishes the core methodology of dimension-reduced cosine expansions, together with the efficient training algorithm, and applies it to netting-set–level MtM and exposure calculations. Among several  tensor decomposition techniques, we focus on Canonical Polyadic Decomposition (CPD) for CCR quantification due to its simplicity and transparency, which
are particularly appealing in practical risk-management applications.

Numerical experiments on netting sets comprising tens of thousands of trades driven by seven risk factors demonstrate that the proposed method achieves relative errors below $0.1\%$ while requiring only a small fraction of the runtime of MC simulation. 

The main computational bottleneck still resides in the offline training stage, where the curse of dimensionality persists. This motivates several follow-up directions, including: (i) replacing CPD by tensor train (TT) decomposition and exploiting recent cross-approximation techniques to cover a substantially higher number of dimensions, particularly useful for margined netting sets; (ii) incorporating instrument-level acceleration techniques developed for MC frameworks to extend the product coverage to including exotic and path-dependent products; and (iii) combining COS--tensor approximations with importance sampling, which both extends the approach from liquid portfolios to general portfolios and enables counterparty-level calculations when a direct COS--tensor evaluation is not feasible. 
\end{abstract}

\section{Introduction}\label{sec: Introduction}

Accurate quantification of counterparty credit risk (CCR) underpins a wide range of financial applications, including valuation adjustments (xVAs), regulatory capital modeling under Basel requirements, and internal risk management practices such as Value-at-Risk (VaR)-based trading limits.

In practice, CCR exposures are almost universally computed using Monte Carlo (MC) simulation. The appeal of MC lies in its flexibility, ease of implementation, and transparency, particularly when risk-factor dynamics do not admit analytically tractable distributions. However, for large portfolios particularly characterized by numerous trades, MC estimators can exhibit substantial variance, particularly for tail quantities such as high-quantile measures of Potential Future Exposure (PFE). Achieving reliable and statistically robust tail estimates therefore necessitates a very large number of simulation paths, which is  computationally prohibitive. In practice, such computational constraints often limit the feasible number of simulations, potentially resulting in statistically noisy and unstable PFE estimates, which may in turn trigger supervisory findings and regulatory risk-mitigation requirements.

\begin{figure}[h!]
\centering
\includegraphics[width=0.5\textwidth]{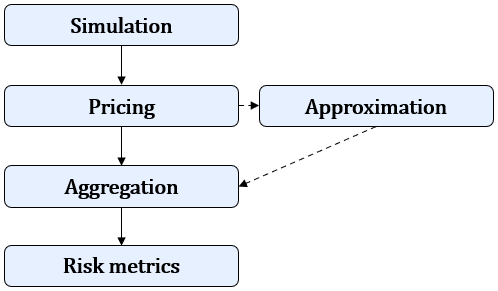}
\caption{Schematic of the Monte Carlo framework for CCR quantification. Solid arrows represent full re-evaluation, while dashed arrows indicate accelerated calculations, for example via Chebyshev interpolation as in \cite{laris2018chebyshev}.}
\label{fig: MC framework representation}
\end{figure}

Most existing research has therefore focused on improving MC efficiency through variance-reduction techniques and instrument-level pricing accelerations, while remaining within the MC paradigm (see Section~\ref{sec: Literature Review}). Figure~\ref{fig: MC framework representation} illustrates the typical MC workflow for CCR calculations.

This paper initiates a series of studies that move beyond Monte Carlo simulation and develop a deterministic \emph{COS--tensor} framework for CCR quantification. The framework is designed for large and liquid portfolios, characterized by a modest number of dominant risk factors and many trades. 

\begin{figure}[h!]
\centering
\includegraphics[width=0.5\textwidth]{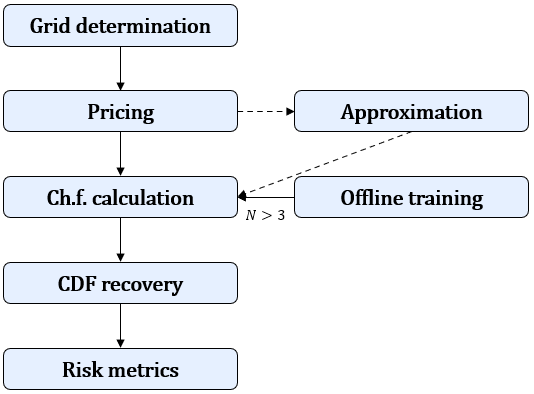}
\caption{The proposed COS--tensor (deterministic) framework for CCR quantification. Solid black lines denote components introduced in this paper, while dashed lines indicate compatibility with existing pricing acceleration techniques. When the number of risk factors exceeds three, the joint density is approximated by a low-rank tensor decomposition obtained through an offline training procedure.}
\label{fig: COS framework representation}
\end{figure}

It rests on three key insights. First, rather than generating exposure scenarios, CCR quantification can be reformulated in the Fourier domain. Once the characteristic functions of portfolio MtM value or exposures is computed numerically, one can recover the corresponding distribution   in a single step via the one-dimensional COS method \cite{fang2009novel}. Second, we derive a \emph{dimension-reduced cosine expansion framework} that effectively mitigates the curse of dimensionality arising in the numerical evaluation of characteristic functions. Specifically, a continuous real-valued function is represented by its Fourier--cosine series, and the resulting coefficient tensor is approximated using a
tensor-decomposition technique. After rearranging terms, this yields a dimension-reduced cosine expansion. When applied to the joint density function appearing in the integration-based definition of netting-set characteristic functions, and by exploiting the separability of netting-set MtM value or exposure, we obtain COS--tensor-based formula for the
corresponding characteristic function. This approach shifts the originally exponential computational complexity to an offline tensor decomposition training stage. Figure~\ref{fig: COS framework representation} illustrates the proposed COS--tensor workflow. Third, the offline tensor decomposition can be done very efficiently by exploiting the  relationship between the joint characteristic function and the Fourier--cosine coefficient tensor following the central insight of the original COS method. That is, we  populate the coefficient tensor directly using the characteristic function and directly train the tensor decomposition in the Fourier domain, which improves both training speed and accuracy by more than two orders compared with conventional gradient-based optimization in the physical domain.

This paper establishes the core methodology of dimension-reduced cosine expansions, together with the associated efficient training algorithm, and their application to netting-set–level mark-to-market (MtM) value and exposure calculations. It also focuses on Canonical Polyadic Decomposition (CPD) for the context of CCR,  due to its transparency and conceptual simplicity, which are particularly appealing in practical
risk-management applications.

In summary, the main contributions of this paper are as follows:
\begin{itemize} \item A portfolio-level CCR framework that reformulates exposure quantification problems in the Fourier domain. \item Derivation of dimension-reduced Fourier--cosine
expansions, using CPD as an example, and an associated efficient Fourier-domain training algorithm for tensor decomposition that is tailored to density functions and significantly improves both accuracy and training efficiency compared to physical-domain optimization.
\item An extensive numerical validation, which demonstrate fast error convergence of the COS--CPD method as its well as stable and superior performance compared with Monte Carlo methods for large liquid portfolios.
\end{itemize}

The main computational bottleneck of the COS--tensor framework lies in the offline training stage, where dimensionality challenges persist. This limitation, together with the deliberately focused scope of the present paper, points to several directions of follow-up research within this series. These include: replacing CPD by tensor train (TT) decomposition and leveraging recent advances in cross approximation for TT formats
\cite{oseledets2009breaking,OseledetsTyrtyshnikov2010} to  reduce high-dimensional training costs, so as to cover  substantially higher dimensions, such as the case for  margined netting sets; incorporating instrument-level acceleration techniques developed for Monte Carlo
frameworks to extend the product coverage to including exotic and path-dependent products; combining COS--tensor approximations with importance sampling: the densities obtained for liquid subportfolios can yield  an effective auxiliary density for importance sampling in the full-portfolio MC simulation, and the same mechanism also enables counterparty-level calculations, in case a direct COS--tensor evaluation is feasible (as explained in Section~\ref{subsec:chf_cpty_exposure}). The main ideas underlying these extensions are outlined in Section~\ref{sec: Extensions} and will be detailed in subsequent papers.

The remainder of the paper is organized as follows.
Section~\ref{sec: Literature Review} reviews related work.
Section~\ref{sec: The modeling framework} introduces the dimension-reduced
cosine expansion based on tensor decomposition.
Section~\ref{sec: algorithm} presents the Fourier-domain CPD training
algorithm.
Section~\ref{sec: COS for CCR} develops the COS--tensor framework for CCR
quantification.
Section~\ref{sec: numerical_tests} reports numerical results.
Section~\ref{sec: Extensions} discusses extensions, and
Section~\ref{sec: Conclusion} concludes.

\section{Literature review}\label{sec: Literature Review}

For clarity, we first provide some basic definitions before surveying related work.

\subsection{Exposure metrics}
\label{subsec: Exposure quantification}

Credit exposure is usually quantified at two hierarchical levels: the \textit{netting-set level} and the \textit{counterparty level}. 

A netting set comprises all transactions covered by a contractual netting
agreement. The exposure of the $j$-th netting set that a bank has with a
counterparty at time $t$ is defined as the positive part of its
Mark--to--Market (MtM) value,
\begin{equation}
E_t^{(j)} := \max\!\bigl( V^{(j)}(\mathbf{Y}^{(j)}_t,t),\, 0 \bigr),
\label{eq:E_nettingset}
\end{equation}
where $\mathbf{Y}^{(j)}_t$ is a vector-valued state process capturing the stochastic drivers  for $j$-th netting set, and $V^{(j)}(\mathbf{Y}^{(j)}_t,t)$ is the total MtM value of the
$j$-th netting set.

For margined netting sets, exposure is typically defined over a margin period of risk (MPOR) of length $\delta>0$. In this case, the relevant exposure at time $t$ is the  change in MtM over the MPOR, floored at zero\footnote{Sometimes a forward-looking definition is also used, i.e., $E^{(j)}_{t,\delta}  :=    \max\!\left(
      V^{(j)}_{t+\delta} - V^{(j)}_{t},\, 0    \right).$, which does not affect the application of this framework.}, i.e.,
\begin{equation}
E^{(j)}_{t,\delta}
:=
\max\!\left(
  V^{(j)}(\mathbf{Y}^{(j)}_{t},t)
  -
  V^{(j)}(\mathbf{Y}^{(j)}_{t-\delta},t-\delta),
  \, 0
\right).
\label{eq:E_nettingset_margined}
\end{equation}
For notational simplicity, the derivations in this paper focus on non-margined netting sets. We note that the methodology developed later for non-margined netting sets extends directly to the margined case, except that for margined netting sets, twice as many risk factors enter, as the MtM values at both $t$ and $t-\delta$ appear in the definition.

For counterparties with multiple netting sets, the counterparty exposure is the sum of the netting-set exposures, i.e.,
\begin{equation}    
\label{eq:E_cpty}
E_t^{\mathrm{cpt}} := \sum_{j=1}^{J}  E_t^{(j)} = \sum_{j=1}^{J} \max \big( V^{(j)}(\mathbf{Y}^{(j)}_t, t), 0 \big),
\end{equation}
where $J$ denotes the number of netting sets between two counterparties. If only a single netting set exists between them, the counterparty exposure coincides with the netting-set exposure.

In the following sections, we drop the superscript 
$(j)$ when focusing on derivations for a single netting set.

Expected Exposure (EE) at a future time point $t$ is a fundamental metric to quantify CCR. It is defined as the  expectation of  $E_t$ given  information available at time $0$ (i.e., the filtration $\mathcal{F}_0$):
\begin{equation}
    \mathrm{EE}(t) = \mathbb{E}\!\left[ E_t \mid \mathcal{F}_0 \right].
    \label{eq: EE general}
\end{equation}
Here, $E_t$ denotes either the netting-set exposure $E_t^{(j)}$ in Eq. (\ref{eq:E_nettingset}) or the counterparty exposure $E_t^{\mathrm{cpt}}$ in Eq. (\ref{eq:E_cpty}), depending on the level of aggregation. Sometimes, discounting is included to yield the discounted version of EE, i.e., 
\(
    \mathrm{EE}_{\text{disc}}(t) = \mathbb{E} \Big[ e^{-\int_0^t \gamma_s \, ds} \, E_t \Big],
\)
with $\gamma$ denoting the  short rate in the same currency as $E_t$.

PFE is a quantile-based metric of the exposure distribution at a future time point $t$. It is effectively the Value-at-Risk of $E_t$: for a given confidence level $\alpha$, 
\begin{equation}\label{eq:PFE}
    \PFE = \inf \big\{ e : \mathbb{P}(E_t \leq e) \geq \alpha \big\},
\end{equation}
where the distribution is usually taken under the \emph{real-world measure}. PFE is widely used in practice to set trading limits at the counterparty level. Some institutions also rely on Expected Shortfall (ES) instead of VaR/PFE.

\begin{remark}
EE is a key input for xVAs, in which case the expectation is taken under the \emph{risk-neutral measure} $\mathbb{Q}$. In practice, however, some institutions also compute the EE profile under the \emph{real-world measure} for comparison with PFE. In this paper, we do not prescribe which measure should be used, but instead assume that model parameters are calibrated under the measure relevant for the application.
\end{remark}

Additionally, EE forms the basis for related quantities such as Expected Positive Exposure (EPE), Expected Negative Exposure (ENE), and Effective Expected Positive Exposure (EEPE). 

In this paper, we focus on PFE and EE. Other metrics can be obtained in a very similar manner.

\subsection{Credit exposure calculation methods in literature}

Simulation-based approaches dominate exposure calculations for their robustness, flexibility, and ease of implementation. Yet their slow convergence makes accurate estimation of high-quantile metrics such as PFE (Eq. \ref{eq:PFE}) computationally demanding, especially for large portfolios. Numerous variance reduction methods have been developed to mitigate this.

Importance sampling shifts the distribution toward rare, high-impact scenarios and thereby reduces variance in tail-risk metrics such as PFE \cite{glasserman1999importance, deng2012sequential}. However, its effectiveness depends on careful calibration and model-specific adjustments, which limit its broader applicability \cite{chatterjee2018sample}. 

Multilevel Monte Carlo (MLMC) \cite{giles2008multilevel} improves efficiency by combining simulations at different discretization levels, while Multi-Index Monte Carlo (MIMC) extends this idea across multiple parameters \cite{hajiali2016multi}. However, both methods encounter scalability issues: MLMC can suffer from diminishing gains in high dimensions, and MIMC often imposes heavy memory requirements and typically relies on smoothness assumptions. 

Quasi-Monte Carlo (QMC) methods substitute pseudo-random numbers with low-discrepancy sequences such as Sobol or Halton points, which accelerates convergence in smooth, low-dimensional settings \cite{glasserman1998quasi}. Their advantage, however, quickly erodes under discontinuities or in high dimensions. Smoothing techniques can mitigate these shortcomings and partially restore efficiency \cite{bayer2023numerical}. 

Functional quantization offers another deterministic alternative by replacing random sampling with Karhunen–Loève–based path expansions \cite{pages2005functional}. Although effective for low-dimensional problems, its complexity generally grows exponentially with the number of risk factors, restricting its practical use.

Several approaches aim to accelerate MC simulation by reducing the cost of pathwise revaluation. Finite Difference Monte Carlo (FDMC) methods pre-compute pricing grids to speed up exposure calculations for European-style contracts \cite{tavella2000pricing, de2014efficient}. Their efficiency, however, breaks down for path-dependent payoffs. Least Squares Monte Carlo (LSMC) handles such payoffs by regressing continuation values on simulated paths \cite{longstaff2001valuing}, but its accuracy depends heavily on the choice of basis functions and can exhibit approximation bias. The Stochastic Grid Bundling Method (SGBM) improves on LSMC by performing local regressions within adaptively clustered paths \cite{karlsson2016counterparty, arregui2024efficient}, and thus, reduces approximation bias. Yet, it still faces dimensionality and basis-selection challenges.  

The Monte Carlo–COS (MCCOS) method \cite{shen2013benchmark} avoids regression altogether by embedding Fourier–cosine expansions into MC simulation, which offers efficient pricing of early-exercise options. Nonetheless, MCCOS, as well as other methods mentioned above, scales poorly in portfolio-level settings, as they evaluate each instrument separately and stay within the MC framework.

Polynomial surrogate models provide another acceleration route. Chebyshev interpolation methods \cite{gass2018chebyshev, glau2019new, glau2021speed} precompute prices on Chebyshev nodes and construct polynomial approximations, enabling fast scenario evaluations, but interpolation complexity grows exponentially with dimensionality. Sparse-grid interpolation methods \cite{grzelak2022sparse} alleviate this burden by selecting collocation points strategically to approximate the portfolio value function, significantly lowering per-path costs. Still, sparse grids expand polynomially with dimension, which makes high-dimensional problems computationally demanding.

Given the scalability limits of pricing function approximations, recent research increasingly turns to machine learning (ML) as an alternative for high-dimensional pricing. Neural networks, in particular, capture complex dependencies and scale efficiently to higher dimensions. Trained offline on simulated or historical data, they allow rapid online evaluation of pricing functions. Applications include the replication and compression of large European option portfolios \cite{dhandapani2024neural}, and the approximation of optimal stopping strategies for Bermudan-style derivatives \cite{andersson2021deep}. Glau et al. \cite{glau2022deep, glau2023neural} show that deep neural networks can solve high-dimensional parametric PDEs for portfolio-level pricing problems in up to ten dimensions. Similarly, neural-network-based methods for backward stochastic differential equations (BSDEs) \cite{han2017deep, gnoatto2023deep} yield accurate valuations and xVA computations for large portfolios.  

Despite these successes, ML methods rely on complex nonlinear transformations and implicit internal representations, which undermines interpretability and reproducibility. This lack of transparency is especially problematic in regulatory risk management, where models must remain explainable, robust, and auditable.


For certain products, computing credit exposures reduces to problems similar to derivative pricing, where deterministic methods can be applied. The Approximate Analytical Solution (AAS), for instance, provides explicit formulas for EE and PFE under the assumption of normally distributed risk factors \cite{li2020explicit, li2022analytical}. While computationally efficient, these formulas rely on restrictive distributional assumptions, which may limit their accuracy over long horizons or in realistic market conditions \cite{natenberg2012option}.  

Fourier-based approaches offer an alternative by reformulating instrument-level exposure computation in the Fourier domain. 
Initially developed for option pricing by Carr and Madan \cite{carr1999option} and subsequently refined by Fang and Oosterlee through the COS method \cite{fang2009novel}, these techniques achieve exponential convergence when payoffs or density functions are smooth. Their application, however, has been largely restricted to single instruments or low-dimensional problems, since evaluating multivariate characteristic functions becomes computationally prohibitive in high dimensions.  

Overall, extending these deterministic methods from individual products to high-dimensional portfolios remains challenging due to the exponential growth in computational complexity with respect to the number of dimensions.

\subsection{Tensor decompositions}

Low-rank tensor decomposition methods have emerged as a powerful tool to reduce the computational complexity of high-dimensional integrals. Unlike (deep) neural networks, tensor decompositions provide structured and transparent approximations of multivariate functions, with theoretically quantifiable error bounds, a desirable feature in regulated financial contexts. Widely used tensor decomposition techniques include Tucker decomposition, Tensor Train (TT) decomposition, and Canonical Polyadic decomposition (CPD). 

Tucker decomposition factorizes a tensor into a lower-dimensional core tensor and mode-specific factor matrices. While it effectively captures interactions among variables, the core tensor remains high-dimensional, and variants such as Hierarchical Tucker alleviate this issue only partially, at the cost of increased implementation complexity \cite{grasedyck2010hierarchical, hackbusch2009new}.

TT decomposition addresses scalability more directly by representing a tensor as a sequence of three-dimensional cores connected in a linear chain \cite{oseledets2011tensor}. This structure achieves linear storage complexity and maintains spectral accuracy, which is advantageous in high-dimensional settings. 

CPD expresses a tensor as a sum of rank-one components, each formed by the outer product of mode-specific vectors \cite{kolda2009tensor}. CPD is often less accurate than TT decomposition for a given storage budget, but simpler in structure.

In this paper, we adopt the CPD for portfolio-level credit exposure calculations, as it provides a good balance of transparency and computational efficiency compared to other tensor decomposition methods.

\subsection{Combination of tensor decomposition and function expansions}
Several prior studies have explored the integration of tensor decomposition with function expansions, though existing approaches differ fundamentally from the framework proposed here.

Glau et al. \cite{glau2020low} combine Chebyshev polynomial interpolation with TT decomposition to approximate option prices as functions of multiple model parameters. High-dimensional Chebyshev interpolants are compressed using the TT format to mitigate the curse of dimensionality. This enables rapid evaluation during the online phase. However, this method operates at the instrument level, i.e., it interpolates the prices of each financial instrument individually per MC path, and thus, serves as a pricing acceleration method within the MC framework.

Bigoni et al. \cite{bigoni2016spectral} introduce a continuous analogue of the discrete TT format --- the functional TT expansion. 
More precisely, they establish a connection between the TT cores of a multivariate function and those of the coefficient tensor of polynomial-based spectral expansion of that function. However, when it comes to the computation of the TT cores, they have to discretize that multivariate function first, so as to apply existing discrete TT decomposition algorithms. This approach is indirect and therefore inefficient. Even though the focus of this paper is not on TT, we will briefly discuss in Section \ref{subsec:CPD-COS_training}, that when replacing polynomial expansions by Fourier-cosine expansion, one can rely on the central insight of the COS method and directly solve the tensor decomposition in the Fourier domain without the need to discretize the target function.

The closest related work combining tensor decomposition with Fourier-series representations is~\cite{Kargas2021}. There, a Fourier Series Approximation based on Hidden Tensor Factorization (FSA--HTF) is proposed for numerical integration. The key observation is that the integration of a function corresponds to the zeroth-order coefficient of its Fourier--cosine expansion. Since the computation of Fourier-series coefficients in high dimensions suffers from the curse of dimensionality, the associated coefficient tensor is approximated using CPD. The CPD, however, is trained in the physical domain by fitting the CPD-based expansion of the target function to sampled function values. This training strategy is indirect and, as shown in Section~\ref{subsec:CPD-COS_training}, exhibits slow convergence and high computational cost. Moreover, no theoretical analysis of the resulting approximation error is provided.

In contrast, this paper develops a general dimension-reduced Fourier--cosine expansion framework, instantiated via CPD and equipped with a highly efficient Fourier-domain training algorithm. 
As for the training method, rather than aggregating the factor matrices into a single function value and fitting it to sampled data, we train the tensor decomposition directly on the Fourier--cosine coefficient tensor by populating its entries via the characteristic function. As demonstrated in Section~\ref{subsec: training performance}, this approach—together with three key algorithmic ingredients—yields substantial improvements in both accuracy and convergence of the training compared with gradient-based optimization in the physical domain. 
\section{Dimension-reduced cosine expansions}
\label{sec: The modeling framework}

This section develops a framework for constructing \emph{dimension-reduced cosine expansions} based on tensor decomposition techniques. While the framework applies to general real-valued continuous functions, the efficient training method developed in this paper exploits characteristic functions and is therefore tailored to probability density functions.

We first review the COS method and then introduce a simple yet effective technique for reducing computational cost in multidimensional settings. Building on this, we derive the main methodological contribution of this paper: a general dimension-reduced cosine expansion framework, with a particular focus on its CPD-based instantiation, referred to as the \emph{CPD--COS expansion}. The associated efficient offline training algorithm tailored to joint densities is presented in Section~\ref{sec: algorithm}.

\subsection{The COS method}\label{subsec: COS method}

Let a finite interval $[a,b]\in\mathbb{R}$ be contained in the support of a density function $f$. $f$ then admits a Fourier--cosine representation on $[a,b]$, i.e.,
\begin{equation}\label{eq:general cosine series}
f(x)
=
\sideset{}{'}\sum_{k=0}^{\infty}
C_k
\cos\!\Big(k\pi \tfrac{x-a}{b-a}\Big),
\qquad x\in[a,b],
\end{equation}
in the $L^2$ sense if $f\in L^2([a,b])$, and uniformly on $[a,b]$ if $f$ is continuously differentiable on $[a,b]$. Uniform convergence is stronger and is typically studied in computational finance applications.
The apostrophe on the summation indicates that the first term is halved.
$C_k$ are the Fourier--cosine coefficients, defined as
\begin{equation}\label{eq:cos coefficients exact}
C_k
:=
\frac{2}{b-a}
\int_a^b
f(x)
\cos\!\Big(k\pi \tfrac{x-a}{b-a}\Big)\,dx.
\end{equation}
Truncating the series after $K$ terms yields the truncated series expansion of $f$:
\begin{equation}
\label{eq:Fourier-cos_pdf}
f(x)
\approx
\sideset{}{'}\sum_{k=0}^{K-1}
C_k
\cos\!\Big(k\pi \tfrac{x-a}{b-a}\Big),
\qquad x\in[a,b].
\end{equation}

The COS method of Fang and Oosterlee \cite{fang2009novel} relates the Fourier--cosine coefficients $C_k$ directly to the characteristic function $\varphi$ of the density. Namely, it holds that
\begin{equation}\label{eq:approximated COS coef}
C_k \approx A_k :=
\frac{2}{b-a}
\operatorname{Re}\!\left\{
\varphi\!\left(\frac{k\pi}{b-a}\right)
e^{-i k \pi a/(b-a)}
\right\},
\qquad k = 0,1,\dots,K-1,
\end{equation} 
if the truncation interval $[a,b]$ is chosen sufficiently large such that the probability mass of $f$ outside $[a,b]$ is negligible. 
Here $\operatorname{Re}\{\cdot\}$ denotes the real part of a complex number. The COS method for density recovery, i.e., 
\[f(x)\approx \hat{f}(x) \quad \text{with}\]
\begin{equation}
\label{eq:1D_COS_pdf}
\hat{f}(x):=
\sideset{}{'}\sum_{k=0}^{K-1}
A_k
\cos\!\Big(k\pi \tfrac{x-a}{b-a}\Big),
\qquad x\in[a,b],
\end{equation}
represents a major breakthrough in computational finance: it avoids numerical Fourier inversion and enables semi‑analytical recovery of densities and option prices using only the characteristic function, which is available for a wide range of jump–diffusion and stochastic volatility models.

The COS method extends naturally to $N$ dimensions. On a sufficiently large domain
$[a_1,b_1]\times\cdots\times[a_N,b_N]$, the joint density admits the approximation
\begin{equation}\label{eq: truncated multivariate Cosine series}
    f(x_1,\cdots, x_N) \approx \hat{f}(x_1,\cdots, x_N):=\sideset{}{'}\sum_{k_1 = 0}^{K_1-1} \cdots \sideset{}{'}\sum_{k_N = 0}^{K_N-1} \mathcal{A}_{\mathbf{k}} \prod_{n=1}^{N} \cos \left(k_n \pi \frac{x_n-a_n}{b_n-a_n}\right),
\end{equation}
where $\mathbf{k} = (k_1, \ldots, k_N) \in \mathbb{N}^N$ is a multi-index and the coefficient tensor $\mathcal{A}_{\mathbf{k}}$ is again approximated by the (joint) characteristic function.  
A compact expression for $\mathcal{A}_{\mathbf{k}}$ is presented in \cite{junike2023multidimensional}. For ease of implementation, here we provide the element-wise calculation formula:
\begin{equation}\label{eq: full_cos_coeff_tensor_elementwise}
    \mathcal{A}_{\mathbf{k}} 
    = \frac{1}{2^{N-1} \prod_{n=1}^{N} L_n} 
    \sum_{\mathbf{s} \in \mathcal{S}} 
    \operatorname{Re}\left\{ 
      \varphi\!\left( 
        \pi\left(\frac{s_1 k_1}{2L_1}, \dots, \frac{s_N k_N}{2L_N}\right) 
      \right) 
      \exp\!\left(-\frac{1}{2}i \pi \sum_{n=1}^N s_n k_n a_n/L_n \right) 
    \right\},
\end{equation}
where $L_n = \frac{b_n - a_n}{2}$ and $\mathbf{s}=(s_1,\cdots,s_N)$ is taken from $\mathcal{S}$, the collection of vectors of length $N$ with each entry being either $1$ or $-1$ but the first coordinate fixed to 1, i.e.,
\[
\mathcal{S} = \{\, (1, s_2, s_3, \dots, s_N) : s_i \in \{-1,1\} \text{ for } i = 2,\dots,N \,\}.
\]
Note that the coefficients $\mathcal{A}_{\mathbf{k}}$ are independent of the spatial variables $x_1,\ldots,x_N$.

On the one hand, the direct evaluation of Eq.~\eqref{eq: truncated multivariate Cosine series}, corresponding to a high-dimensional tensor--vector product, suffers from the curse of dimensionality. As a result, applications of the COS method in literature have largely been confined to low-dimensional settings. We will address this limitation by the COS-CPD method in Section~\ref{subsec: COS-CPD}.

On the other hand, since $|\mathcal{S}|=2^{N-1}$, the computational cost of evaluating the coefficient tensor grows exponentially with $N$. Although this remains a bottleneck for the COS--CPD method, it can be  alleviated considerably by incorporating recent advances in tensor train decomposition, as outlined in Section~\ref{sec: Extensions}, which we will also detail in a follow-up paper.

\subsection{Enhanced multi-dimensional version: Principal--COS method}
\label{subsec:principal-COS}
Before presenting the main solution, we introduce a simple yet very effective technique that can considerably reduce the computational cost of the multi-dimensional COS method. 

The key insight is that, for smooth density functions, a large proportion of the Fourier–cosine coefficients are negligible in magnitude. The corresponding frequencies therefore contribute little to the density reconstruction and can be safely omitted from the cosine expansion. 
We refer to the remaining significant frequencies as \emph{principal frequencies}. 

A limitation of this observation is that the importance of a frequency can only be determined after the associated cosine coefficient is computed, which then does not really help with reducing the computational time in one-dimensional setting. 

However, in the multidimensional case, we assume that frequencies identified as insignificant
in marginal density reconstructions also contribute negligibly to the joint density.
In other words, the summations in~\eqref{eq: truncated multivariate Cosine series} can be restricted to sets of principal frequencies identified via marginal distributions per dimension. This yields a variant of the multidimensional COS formula, namely,
\begin{equation}\label{eq: multi_principal_cos}
    f(x_1,\ldots,x_N) \approx \hat{f}_{\mathcal{K}}(x_1,\ldots,x_N) :=
    \sideset{}{'}\sum_{k_1 \in \mathcal{K}_1} 
    \cdots 
    \sideset{}{'}\sum_{k_N \in \mathcal{K}_N} 
    \mathcal{A}_{\mathbf{k}}
    \prod_{n=1}^{N} 
    \cos\!\left(k_n \pi \tfrac{x_n-a_n}{b_n-a_n}\right),
\end{equation}
where the set $\mathcal{K}_n$ is determined by retaining only those Fourier--cosine coefficients of the marginal density in dimension $n$ whose magnitudes exceed a prescribed threshold (e.g., $10^{-15}$). We refer to Eq.~\eqref{eq: multi_principal_cos} as the \emph{principal--COS} method, to distinguish it from the COS method employing spectral filtering, as used in Section~\ref{subsec: CDF for Counterparty}.

\begin{figure}[H]
    \centering
    \begin{minipage}[t]{0.49\textwidth}
        \centering
        \includegraphics[width=\textwidth]{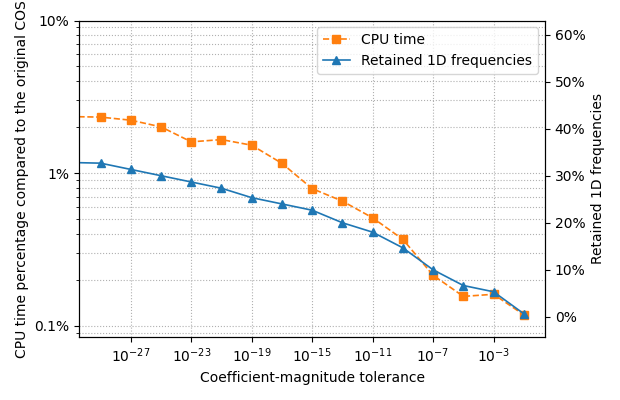}
    \end{minipage}
    \hfill
    \begin{minipage}[t]{0.49\textwidth}
        \centering
        \includegraphics[width=\textwidth]{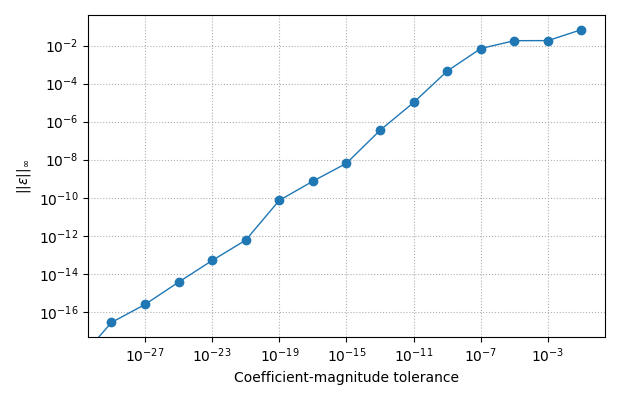}
    \end{minipage}
\caption{Efficiency and accuracy of the principal--COS method. 
Left: CPU time percentage compared to the original COS formulation and fraction of retained principal frequencies (in percentage of the full set of frequencies) versus the coefficient-magnitude tolerance.
Right: approximation error $\|\epsilon\|_\infty$ for the recovered three-dimensional Gaussian density with correlations up to $60\%$. The full COS expansion uses $150$ terms per dimension.}
    \label{fig:principal_cos_demo}
\end{figure}

Numerical experiments show that the principal--COS method significantly reduces runtime relative to the original COS formulation, while incurring negligible loss of accuracy, even in the presence of strong correlations. As an illustration, Fig.~\ref{fig:principal_cos_demo} presents results for a three-dimensional standard Gaussian density with correlations up to $60\%$. The left panel displays the CPU time of the principal--COS method as a percentage of that of the original COS formulation (left vertical axis), together with the fraction of retained principal frequencies relative to the full set of $150^3$ cosine terms (right vertical axis). It can be seen that the CPU time is substantially reduced, since more than half of the frequencies can be discarded. This computational gain is expected to be even more pronounced in higher dimensions. The right panel shows the approximation error, measured in the infinity norm $\|\epsilon\|_\infty$, as a function of the coefficient-magnitude threshold. The approximation error increases gradually as the coefficient-magnitude tolerance is relaxed. At a tolerance of $10^{-19}$, the principal–COS method requires only about $1\%$ of the CPU time of the full COS expansion while the approximation error remains below $10^{-10}$, demonstrating substantial computational savings with limited loss of accuracy.

Principal--COS filtering is also crucial for the offline training stage, as it
substantially improves the convergence speed of the tensor decomposition;
see Section~\ref{sec: algorithm}.

\subsection{Dimension-reduced cosine expansions and the COS--CPD instantiation}\label{subsec: COS-CPD}

Although Eq.~\eqref{eq: multi_principal_cos} substantially reduces computational cost,
its complexity remains exponential in $N$. To tackle this curse of dimensionality, we derive \emph{dimension-reduced cosine expansions} by applying tensor decomposition techniques to the coefficient tensor $\mathcal{A}_{\mathbf{k}}$.

In the context of CCR quantification, we focus on CPD due to its structural simplicity and interpretability. We refer to the resulting instantiation as the \emph{COS--CPD expansion}.

\begin{remark}
Other tensor decomposition techniques are also applicable within the proposed framework.
In particular, a COS--TT expansion based on tensor train (TT) decomposition, including full implementation and comprehensive numerical validation, has been completed and will be reported in a forthcoming paper. Compared with CPD, TT decomposition employs three-dimensional core tensors
and can therefore achieve higher approximation accuracy. Moreover, recent advances in efficient TT approximation algorithms can substantially extend the range of dimensions. More details are given in Section~\ref{subsec:TT-COS}.
\end{remark}

Let us first recall the general CPD representation. A tensor with $N$ dimensions is referred to as  a $N$-th order tensor. The CPD rank of a tensor, $R$, is the minimum number of rank-1 tensors needed to produce this tensor as their sum. That is, an $N$-th order tensor $\mathcal{A} \in \mathbb{R}^{K_1 \times \dots \times K_N}$ can be expressed as a sum of $R$ sets of outer products of $N$ vectors, i.e.,
\begin{equation*}
    \mathcal{A} = \sum_{r=1}^{R} a_r^1 \circ a_r^2 \circ \dots \circ a_r^N,
\end{equation*}
where $a_r^n \in \mathbb{R}^{K_n}$ is called the mode-$n$ factor vector and $\circ$ denotes the outer product. 
The above is equivalent to the following element-wise calculation:
\begin{equation}
    \mathcal{A}[k_1, k_2, \dots, k_N] = \sum_{r=1}^{R} a_r^1[k_1] \, a_r^2[k_2] \, \dots \, a_r^N[k_N].
\end{equation}
Collecting the rank-1 components into matrices, we have
\begin{equation}\label{eq:cpd_coeff}
    \mathcal{A}[k_1, k_2, \dots, k_N] = \sum_{r=1}^{R} \mathbf{A}_1[k_1,r] \, \mathbf{A}_2[k_2,r] \, \dots \, \mathbf{A}_N[k_N,r] = \Bigg(\sum_{r=1}^{R} \prod_{n=1}^N \mathbf{A}_n[k_n,r]\Bigg),
\end{equation}
where the matrices are called factor matrices, i.e, $
    \mathbf{A}_n = 
    \begin{bmatrix}
        a_1^n & a_2^n & \dots & a_R^n
    \end{bmatrix} 
    \in \mathbb{R}^{K_n \times R}.
$

Unlike the matrix setting, where the Eckart--Young--Mirsky theorem guarantees optimal low-rank approximations via the singular value decomposition (SVD), the situation for tensors is more intricate. For tensors of order three or higher, best rank-$R$ approximations (in the CP-rank sense) may fail to exist. Even when they do, such approximations can be non-unique and ill-posed~\cite{de2008tensor}. Furthermore, determining the exact tensor rank is NP-hard~\cite{Hastad1990}. In practice, one therefore computes approximate low-rank decompositions with a prescribed target rank $\tilde R$.

Next, we apply CPD  (with rank $\tilde R$) to the tensor of Fourier--cosine coefficients $\mathcal{A}_{\mathbf{k}}$ in Eq.~\eqref{eq: multi_principal_cos}.
After  replace the coefficient tensor by its CPD approximation, we interchange the order of summations and merge the multiplications along the same dimensions to yield
\begin{align}
    f(x_1,\dots,x_N) &\approx 
     f_{\mathrm{cpd}}(x_1,\ldots,x_N) \quad \text{with}\nonumber\\
     f_{\mathrm{cpd}}(x_1,\ldots,x_N) &:=
    \sum_{k_1\in\mathcal{K}_1} \cdots \sum_{k_N\in\mathcal{K}_N}
    \Bigg(\sum_{r=1}^{\tilde{R}} \prod_{n=1}^N \mathbf{A}_n[k_n,r]\Bigg) 
    \prod_{n=1}^N \cos\!\left(k_n \pi \tfrac{x_n-a_n}{b_n-a_n}\right) \nonumber \\
    &= \sum_{r=1}^{\tilde{R}} \prod_{n=1}^N \Bigg( \sum_{k_n\in\mathcal{K}_n} \mathbf{A}_n[k_n,r] \, \cos\!\left(k_n \pi \tfrac{x_n-a_n}{b_n-a_n}\right) \Bigg).\label{eq:COS-CPD_ele}
\end{align}
The prime notation in Eq.~\eqref{eq: multi_principal_cos}, which halves the $k_n=0$ terms, is omitted here, as these weights are absorbed into the corresponding rows of the factor matrices~$\mathbf{A}_n$. Moreover, for each dimension $n$ we define the basis vector consisting only of the retained frequencies as
\[
\mathbf{v}_n(x_n)
:=
\bigl\{
\cos\!\bigl(k_n \pi \tfrac{x_n-a_n}{b_n-a_n}\bigr)
\bigr\}_{k_n \in \mathcal{K}_n}
\in \mathbb{R}^{|\mathcal{K}_n|\times 1 }.
\]
With this notation, each inner summation in~\eqref{eq:COS-CPD_ele} can be written compactly as
$\mathbf{v}_n(x_n)^\top \mathbf{A}_n[:,r]$, which gives
\begin{equation}\label{eq:CPD form PDF}
{f}_{\mathrm{cpd}}(x_1,\ldots,x_N)=
\sum_{r=1}^{\tilde{R}}
\prod_{n=1}^{N}
\mathbf{v}_n(x_n)^\top \mathbf{A}_n[:,r].
\end{equation}

We emphasize that an analogous procedure extends to other tensor‑decomposition formats; CPD is used here simply as a concrete and transparent instance of the proposed framework.

If the same number of cosine terms $K$ is used across all dimensions, the computational cost of evaluating a single density point reduces to $O(\tilde{R}NK)$  using Eq.~\eqref{eq:CPD form PDF}, instead of $O(K^N)$ as using Eq.~\eqref{eq: truncated multivariate Cosine series}. Thus, the curse of dimensionality is effectively mitigated in the (on-line) evaluation of the joint density function.

The (offline) training of the factor matrices $\mathbf{A}_n$ is discussed in Section \ref{sec: algorithm}. This step involves one additional error as the factor matrices are very close to but are not identical to the true ones. We denote the final formula as 
\begin{equation}\label{eq:COS-CPD-PDF}
\tilde{f}_{\mathrm{cpd}}(x_1,\ldots,x_N):=
\sum_{r=1}^{\tilde{R}}
\prod_{n=1}^{N}
\mathbf{v}_n(x_n)^\top \tilde{\mathbf{A}}_n[:,r].
\end{equation}
We refer to Eq.~\eqref{eq:COS-CPD-PDF} as the \emph{CPD-based dimension-reduced cosine expansion}, or simply the \emph{CPD--COS expansion}, of the joint density function.


\section{Associated training algorithm}\label{sec: algorithm}

In this section, we develop the associated  training algorithm for the COS--CPD expansion. To highlight the distinction between the proposed training algorithm and existing methods, we first briefly review the most relevant method in the literature and adopt it as a benchmark.


\subsection{Benchmark: Physical-domain training via alternating least squares}
\label{subsec:physical_training}

As discussed in Section~\ref{sec: Literature Review}, the most related work in the literature that applies CPD to Fourier-series coefficient tensors is~\cite{Kargas2021}. In that work, the CPD model is trained in the physical domain using gradient-based optimization.
More precisely, since computing individual Fourier coefficients of a target function $f$ from their defining integrals is computationally expensive, an indirect strategy is adopted.

To distinguish the benchmark method from our proposed approach, we denote its computed CPD factor matrices by $\hat{\mathbf{A}}_n$.

First, samples $\{f_m\}_{m=1}^M$ of the target function are collected, where $f_m = f(\mathbf{x}^m)$. The approximate CPD factor matrices $\{\hat{\mathbf{A}}_n\}_{n=1}^N$, with
$\hat{\mathbf{A}}_n \in \mathbb{R}^{I_n \times \tilde R}$, are then obtained by minimizing the squared discrepancy between the sampled function values and their CPD-based Fourier-series approximation. Here, $I_n:= K_n-1$ is the number of Fourier--cosine expansion terms in mode $n$.
Specifically, the following optimization problem is considered:
\begin{equation}
\label{eq: optimization real}
\min_{\{\hat{\mathbf{A}}_n\}_{n=1}^N}
\frac{1}{M}
\sum_{m=1}^M
\left(
f_m -
\left(
\circledast_{n=1}^N
\bigl(
\mathbf{v}_n[:,m]^{\!\top}\hat{\mathbf{A}}_n
\bigr)
\right)\mathbf{1}
\right)^2,
\end{equation}
where $\circledast$ denotes the Hadamard (element-wise) product across modes, and the vectors $\mathbf{v}_n[:,m]$ and factor matrices $\hat{\mathbf{A}}_n$ follow the notation introduced in
Subsection~\ref{subsec: COS-CPD}. The rank parameter $\tilde R$ denotes the chosen approximation rank, which may differ from the true CP rank of the underlying Fourier coefficient tensor.

Due to the multilinear structure of the CPD model, the optimization problem \eqref{eq: optimization real} is typically solved using an Alternating Least Squares (ALS) strategy. ALS proceeds by iteratively updating one factor matrix at a time while keeping the remaining factor matrices fixed, cycling through all modes until a prescribed tolerance is reached or a maximum number of
iterations is exceeded. As summarized in \cite{kolda2009tensor,grasedyck2013literature}, numerous algorithms exist for computing CPD with a fixed rank; nevertheless, ALS and its variants remain the
standard workhorse. The origins of ALS for CPD can be traced back to \cite{CarrollChang1970}.

When all factor matrices except one, say $\hat{\mathbf{A}}_n$, are fixed, problem~\eqref{eq: optimization real} reduces to the following single-mode subproblem:
\begin{equation}
\label{eq: optimization ALS real}
\min_{\hat{\mathbf{A}}_n \in \mathbb{R}^{I_n \times \tilde R}}
\frac{1}{M}
\sum_{m=1}^M
\left(
f_m -
\mathbf{v}_n[:,m]^{\!\top}
\hat{\mathbf{A}}_n
\mathbf{Q}_n[:,m]
\right)^2,
\end{equation}
where
\[
\mathbf{Q}_n[:,m]
:=
\circledast_{j \ne n}
\bigl(
\hat{\mathbf{A}}_j^{\!\top}\mathbf{v}_j[:,m]
\bigr)
\in \mathbb{R}^{\tilde R}
\]
denotes the mode-wise Hadamard product of the contributions from the remaining factor matrices evaluated at the sample point $\mathbf{x}^m$. This subproblem admits an explicit gradient with respect to $\hat{\mathbf{A}}_n$ and is therefore solved using gradient-based updates within each ALS step.

This physical-domain training approach has two notable limitations:
\begin{enumerate}
\item all information about the Fourier--cosine coefficients is aggregated into scalar function values in the objective, which is methodologically inefficient and typically leads to slow convergence;
\item it requires analytical tractability of the target function $f$, which may be unavailable in practice.
\end{enumerate}

\subsection{Our solution: training in the Fourier-domain}
\label{subsec:CPD-COS_training}
We note that in many real-world applications, including CCR quantification, the core computational task involves evaluating multi-dimensional expectations. In such settings, the primary bottleneck lies in handling the joint density function, in particular its lack of separability. Moreover,
the joint density is often unavailable in closed form, whereas the corresponding characteristic function is.

By exploiting the central insight of the COS method, the Fourier--cosine coefficient tensor can be expressed directly in terms of the characteristic function. This enables tensor decompositions, such as CPD, to be evaluated directly in the Fourier domain. 
That is, the cosine coefficient tensor (of the joint density function of all state variables) is first approximately by ${\mathcal{A}}_{\mathbf{k}}$, obtained by inserting the characteristic function into \eqref{eq: full_cos_coeff_tensor_elementwise}.

We then apply ALS  to compute the factor matrices. At each ALS sweep, one factor matrix is updated while the remaining factor matrices are kept fixed. For mode $n$, we consider a transposed
least-squares formulation and compute $\mathbf{A}_n$ by solving
\begin{equation}
\mathbf{A}_n \leftarrow
\arg\min_{\mathbf{A}_n \in \mathbb{R}^{|\mathcal{K}_n| \times R}}
\left\|
\mathbf{Q}_n \mathbf{A}_n^{\top} - \mathbf{B}^{(n)}
\right\|_{\mathrm{F}}^{2},
\label{eq:als_subproblem}
\end{equation}
where
\[
\mathbf{Q}_n :=
\bigodot_{j \neq n} \mathbf{A}_j
\in
\mathbb{R}^{\left(\prod_{j \neq n} |\mathcal{K}_j|\right) \times \Tilde R}
\]
is the Khatri--Rao product of all factor matrices except $\mathbf{A}_n$, and
\[
\mathbf{B}^{(n)} \in
\mathbb{R}^{\left(\prod_{j \neq n} |\mathcal{K}_j|\right) \times |\mathcal{K}_n|}
\]
denotes the transpose of the mode-$n$ matricization of
$\mathcal{A}_{\mathbf{k}}$.
The Frobenius norm is defined as
$\|\mathbf{P}\|_{\mathrm{F}}^{2} = \sum_{i,j} |P_{ij}|^{2}$.


One ALS sweep consists of solving \eqref{eq:als_subproblem} sequentially for $n = 1, \dots, N$; the sweeps are repeated until a prescribed stopping criterion is satisfied, such as the infinity norm of the sampled reconstruction error falling below a prescribed tolerance.

In addition to the main idea outlined above, three further ingredients are essential for the performance:
(i) optimization over randomly selected samples with on-the-fly evaluation, which avoids the explicit storage of the full coefficient tensor; 
(ii) the principal--COS filtering introduced in Section~\ref{subsec:principal-COS}; and
(iii) the use of frequency-dependent weights to prioritize dominant low-frequency components during the early iterations of the training procedure.
These ingredients are detailed in the subsections to follow.

The complete algorithm is summarized in the pseudocode below.

\begin{algorithm}[H]
\caption{Our training method for COS--CPD expansions on joint densities}
\label{algo:our_training_method}
\begin{algorithmic}[1]
\Function{Train}{$\varphi, \Tilde R, M, i_{\max}$}
    \State Determine index sets $\{\mathcal{K}_n\}_{n=1}^N$ \Comment{Principal--COS frequency filtering}
    \State Initialize factor matrices $\mathbf{A}_1, \ldots, \mathbf{A}_N$
    with $\mathbf{A}_n \in \mathbb{R}^{|\mathcal{K}_n| \times \Tilde R}$
    \For{$i = 1, \ldots, i_{\max}$} \Comment{ALS outer loop}
        \State Compute sampling weights $w_i$
        \Comment{Frequency-weighted sampling}
        \For{$n = 1, \ldots, N$} \Comment{Mode-$n$ update}
            \State Define sampling operator
            $\mathcal{S} \in \mathbb{N}^{M \times \prod_{j \neq n} |\mathcal{K}_j|}$
            \State $\mathbf{Q}_S \gets \textsc{SKR}\!\left(
            \mathcal{S},
            \{\mathbf{A}_j\}_{j \neq n}
            \right)$
            \Comment{Sampled Khatri--Rao product}
            \State $\mathbf{B}^{(n)}_S \gets
            \texttt{EvaluateOnTheFly}(\mathcal{S}, \varphi)$
            \Comment{Populate $\mathcal{A}_{\mathbf{k}}$ via Eq.~\eqref{eq: full_cos_coeff_tensor_elementwise}}
            \State $\mathbf{A}_n \gets
            \arg\min_{\mathbf{A} \in \mathbb{R}^{|\mathcal{K}_n| \times R}}
            \| \mathbf{Q}_S \mathbf{A}^{\top} - \mathbf{B}^{(n)}_S \|_{\mathrm{F}}^{2}$
        \EndFor
        \If{termination criteria met}
            \State \textbf{break}
        \EndIf
    \EndFor
    \State \Return factor matrices $\{\mathbf{A}_n\}_{n=1}^N$
\EndFunction
\end{algorithmic}
\end{algorithm}
The inputs to Algorithm~\ref{algo:our_training_method} are the characteristic function $\varphi$, the target CP rank $\Tilde R$, the batch size $M$ (number of sampled mode-$n$ fibers per iteration), and the maximum number of ALS iterations $i_{\max}$. The quantity $|\mathcal{K}_n|$ denotes the number of retained cosine-frequency components in mode $n$ after principal--COS frequency filtering. Each factor matrix
$\mathbf{A}_n \in \mathbb{R}^{|\mathcal{K}_n| \times \Tilde R}$ represents the mode-$n$ factor of the CPD of the Fourier--cosine coefficient tensor $\mathcal{A}_{\mathbf{k}}$. At each ALS iteration, the operator \textsc{SKR} constructs a sampled Khatri--Rao product of all factor matrices except the one currently being updated, while the routine \texttt{EvaluateOnTheFly} computes the corresponding sampled mode-$n$ fibers of $\mathcal{A}_{\mathbf{k}}$ directly from the characteristic function
$\varphi$. The resulting sampled least-squares subproblem is then solved to update the current factor matrix. The ALS iterations are repeated until a prescribed termination criterion is satisfied or the maximum number of iterations $i_{\max}$ is reached.

The  strong performance of this algorithm is demonstrated in Section~\ref{subsec: training performance}.

\subsubsection{Random sampling}
In high-dimensional settings, forming and storing the full coefficient tensor $\mathcal{A}_{\mathbf{k}}$ is computationally infeasible. To address this, we solve the least-squares problem using a randomized algorithm that evaluates only a subset of tensor entries. More precisely, each mode update is obtained by solving a subsampled least squares problem using mode-$n$ fibers and the Khatri–Rao product of the other factors.

To construct the sampled least squares system, the operator \textsc{SKR} (Sampled Khatri–Rao) computes only the required rows of the full Khatri–Rao product of all factor matrices, avoiding full materialization. The resulting matrix $\mathbf{Q}_S$ serves as the design matrix in the least squares problem. Meanwhile, the sampled tensor entries $\mathbf{B}^{(n)}_S$ are evaluated on-the-fly via Eq.~\eqref{eq: full_cos_coeff_tensor_elementwise} using the characteristic function $\varphi$, as implemented in line 8. This avoids storing the full tensor $\mathcal{A}_{\mathbf{k}} \in \mathbb{R}^{K^N}$ and enables scalable training. The resulting overdetermined least squares problem is solved using QR decomposition.


\subsubsection{Principal--COS filtering}
\begin{figure}[h!]
    \centering
    \includegraphics[width=0.7\linewidth]{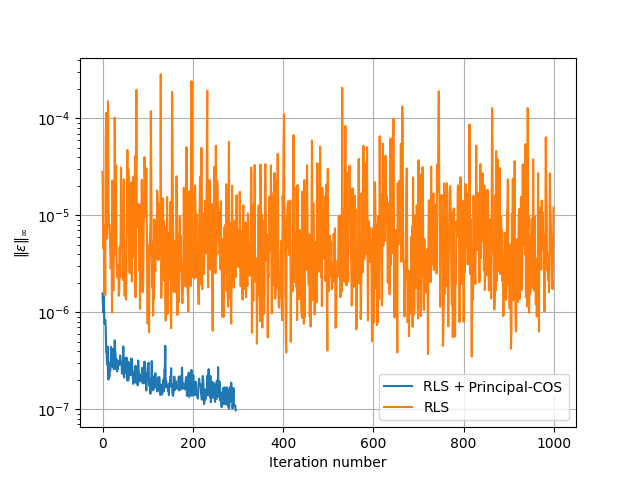}
    \caption{Convergence behavior of Algorithm \ref{algo:our_training_method}  with and without principal--COS filtering. The error is measured in absolute terms using the infinity norm, $|\epsilon|_\infty$, across iterations in a three-dimensional setting.}
    \label{fig: comparison principle cos}
\end{figure}

As a pre-processing step (line 2 of the pseudocode), we apply the frequency-filtering procedure described in Section~\ref{subsec:principal-COS}. For each risk factor, marginal cosine coefficients $A_k$ are computed and frequency indices with coefficients below a fixed threshold (e.g., $10^{-15}$) are discarded. This reduces the number of retained frequencies without compromising accuracy, yielding smaller factor matrices and improved numerical stability in the subsequent tensor-decomposition training. 
Figure~\ref{fig: comparison principle cos} illustrates the approximation error in the infinity norm of the CPD representation of the full coefficient tensor. Within each ALS step, a randomized least-squares (RLS) solver is used to solve the minimization problem in Eq.~\eqref{eq:als_subproblem}. When combined with principal–-COS filtering, this results in faster and more stable convergence of the RLS algorithm.

\subsubsection{Frequency-based weighting}
As shown in line 5 of the pseudocode, we introduce frequency-weighted sampling, deviating from the uniform sampling scheme as commonly used in literature. This modification is motivated by the spectral properties of smooth functions: 
the magnitudes of their cosine series coefficients decay exponentially with increasing frequency. Consequently, the dominant signal energy concentrates in low-frequency components, while high-frequency coefficients approach zero and become numerically challenging to resolve accurately. Assigning uniform weights to the sampled points would introduce unnecessary approximation error and instability. We address this issue by frequency-dependent weighting, 
i.e., assigning higher weights to spectral regions most relevant for reconstruction accuracy.
Within each ALS iteration $i$, we apply a power-law frequency weighting to the sampled frequencies. Specifically, for the $k$-th frequency we define
\begin{equation*}
    w_{k,i} = \frac{1}{(k + 1)^{\alpha_i}}, \quad \text{with normalization} \quad \sum_{k\in\mathcal{K}} w_{k,i} = 1,
\end{equation*}
where $\alpha_i$ decreases over iterations, gradually shifting the sampling distribution from a low-frequency bias toward a more uniform spread. This annealing strategy prioritizes dominant components in early training while enabling finer-scale refinement later on. Among monotone decay schemes, the power-law form is particularly well suited for discrete frequency indices, as it avoids introducing an explicit frequency scale and yields a stable, gradual reweighting across the spectrum.

These weights are used to construct the sampling operator $\mathcal{S}$ in line 7, which selects $M$ mode-$n$ fibers for each dimension $n$.

\begin{remark}
Alternative annealing schemes, such as exponential decay or inverse-time decay, are tested to exhibit similar behavior in practice when their hyper-parameters are appropriately tuned.
\end{remark}

\section{The COS-tensor framework for credit exposures}
\label{sec: COS for CCR}
Building on the dimension-reduced cosine expansions developed earlier, this section constructs a COS--tensor framework for credit exposure quantification in large, liquid portfolios characterized by a modest number of dominant risk factors and a large number of trades.

As outlined in Section~\ref{sec: Introduction}, CCR quantification is reformulated in the Fourier domain. Instead of generating exposure scenarios via Monte Carlo simulation, the characteristic function of the total exposure is evaluated numerically, and the one-dimensional COS method is applied to recover the CDF in a single step. By exploiting the dimension-reduced cosine expansions introduced in Section~\ref{sec: The modeling framework}, and the fact that the MtM value of a liquid portfolio can typically be decomposed into low-dimensional contract legs, the overall computational complexity is reduced to that of low-dimensional numerical integration.

\subsection{Portfolio setup for method illustration}\label{subsec: portfolio setup}

We consider a representative, liquid portfolio composed of linear interest rate (IR) and foreign exchange (FX) derivatives. The associated risk factors are collected in a \((2N+1)\)-dimensional vector process
\[
\mathrm{Z}_t
=
\big[\gamma_{0,t}, \gamma_{1,t}, \ldots, \gamma_{N,t}, X_{1,t}, \ldots, X_{N,t}\big],
\]
where \(\gamma_{0,t}\) denotes the domestic short rate, \(\gamma_{n,t}\), \(1 \le n \le N\), denote foreign short rates, and \(X_{n,t}\), \(1 \le n \le N\), are the corresponding FX spot rates, quoted in units of domestic currency per unit of foreign currency.

For notational simplicity, we assume that each component \(\mathrm{Z}_{i,t}\), \(1 \le i \le 2N+1\), evolves as a one-dimensional continuous-time Markov process of the form
\[
\mathrm{d}\mathrm{Z}_{i,t}
=
\mu_i(t,\mathrm{Z}_{i,t})\,\mathrm{d}t
+
\sigma_i(t,\mathrm{Z}_{i,t})\,\mathrm{d}W_{i,t},
\]
where \(\{W_{i,t}\}_{i=1}^{2N+1}\) are standard Brownian motions with instantaneous correlations
\[
\mathrm{d}W_{i,t}\,\mathrm{d}W_{j,t} = \rho_{i,j}\,\mathrm{d}t.
\]

Each risk factor \(\mathrm{Z}_{i,t}\) is assumed to be a deterministic function of a corresponding state variable \(\mathrm{Y}_{i,t}\), namely
\begin{equation}
\label{eq:mapping_state_variable}
\mathrm{Z}_{i,t} = \phi_i(\mathrm{Y}_{i,t}, t),
\end{equation}
for some deterministic mapping \(\phi_i\). 

The state vector \(\mathrm{Y}_t\) thus contains the minimal stochastic information required to reconstruct all risk factors and is Markovian by construction. Within the COS–tensor framework, we assume that either the joint density or the joint characteristic function of \(\mathrm{Y}_t\) is available in closed or semi-closed form.

This framework encompasses most interest rate and FX models commonly used in practice, including the Hull--White model, the Cox--Ingersoll--Ross (CIR) model, the shifted CIR model, and the geometric Brownian motion (GBM). Moreover, the setting readily extends to multifactor specifications and affine stochastic volatility models by augmenting the dimension of \(\mathrm{Z}_t\) and \(\mathrm{Y}_t\) and adjusting the mappings \(\phi_i\) accordingly.

For portfolios of linear IR and FX instruments, each contract leg depends on at most one interest rate and, possibly, one FX rate. This structure allows us to aggregate contract legs sharing common risk factors into subportfolios, each involving no more than two factors. As a result, the total MtM value of the portfolio, expressed in domestic currency, admits the decomposition
\begin{equation}
\label{eq:portfolio_split}
    V(\mathrm{Z}_t, t)
    =
    V_0\big(\gamma_{0, t}, t\big)
    +
    \sum_{n=1}^{N}
    V_n\big(\gamma_{n, t}, X_{n,t}, t\big),
\end{equation}
where \(V_0\) denotes the value of the subportfolio depending solely on the domestic interest rate, and each \(V_n\) represents the value of a subportfolio traded or settled in foreign currency \(n\), converted into domestic currency via the FX rate \(X_n\).

For calculating CCR, which is done under the real world measure, equation ~\eqref{eq:portfolio_split} together with our assumption ~\eqref{eq:mapping_state_variable} clearly indicates each \(V_n\) depends on its own state variables, although state variables in could be correlated.  

On the other hand, xVAs, e.g., Credit valuation adjustment (CVA), is required to be calculated under the risk neutral measure. Equation ~\eqref{eq:portfolio_split} remains valid, only that in this case \(X_n(t)\) typically involves the difference between \(\int_0^T r_n(u)\,du \) and \(\int_0^T r_0(u)\,du\). 

E.g., if we assume the GBM model for the FX rate \(X_n\) , 

\begin{equation}
X_{n,t}
=
X_{n, 0} \exp\!\left(
\int_0^t \Bigl(r_0(u)-r_n(u)-\tfrac12\sigma_{X_n}^2\Bigr)\,du
+
\sigma_{X_n} W_{X_n, t}
\right),
\end{equation}
where \(\sigma_{X_n}\) denotes the lognormal volatility and \(W_{X_n, t}\) denotes the Brown motion that drives \(X_n\). The drift differential appears. The same holds for time-dependent deterministic FX volatility, local or stochastic volatility.

\begin{remark}
While this illustration focuses on linear IR and FX products, the COS-tensor framework also accommodates derivatives with closed-form pricing formulas, such as IR caps, floors, and European-style FX options. As we will indicate in Section~\ref{sec: Extensions}, a range of exotic derivatives can also be incorporated via existing pricing approximation methods such as those based on Chebyshev polynomial interpolation.
\end{remark}

\begin{remark}
Although the COS-tensor framework is derived primarily for liquid portfolios as described above, it can potentially be useful to more general portfolio settings. In practice, liquid products often account for more than $80\%$ of a portfolio’s total sensitivities or MtM value. As a result, a fast CCR evaluation of the liquid subportfolio using the COS-tensor framework can provide an efficient baseline for the CCR assessment of the full portfolio. For example, as indicated in Section~\ref{subsec:COSIS}, one of our ongoing works investigates the use of the resulting liquid-subportfolio density as an auxiliary density for importance sampling in the evaluation of the total portfolio. 
\end{remark}

\subsection{Computation of characteristic functions using COS-CPD}
\label{subsec: chf calculation}

Throughout this subsection, we fix a future time point $t$ and suppress the time index whenever no confusion arises.

\subsubsection{The characteristic function of portfolio-level MtM value}

Let $\mathbf{Y}$ again denote the vector of state variables — either the driving
random variables of the underlying risk factors or the risk factors themselves — that determine portfolio values, with joint probability density function $f(\mathbf{y})$. 
 
As in Eq.~\eqref{eq:E_nettingset} and~\eqref{eq:E_cpty}, we denote the MtM value of a portfolio by $V$ and the exposure by $E$.

The characteristic function of the MtM value is defined as
\begin{equation}
\label{eq: chf MtM}
\varphi^{V}(\omega)
=
\mathbb{E}\!\left[e^{i\omega V(\mathbf{Y})}\right]
=
\int_{\mathbb{R}^{|\mathbf{Y}|}}
\exp\!\left(i\omega V(\mathbf{y})\right)
f(\mathbf{y})\,d\mathbf{y}.
\end{equation}

In special cases, such as a single European option under affine models, the characteristic function $\varphi^{V}$ admits a closed-form expression; see, for example, \cite{fang2009novel}. For general portfolios, however, closed-form solutions are typically unavailable and $\varphi^{V}$ must be evaluated numerically.

When the number of risk factors is low (e.g., three) and the joint density function $f(\mathbf{y})$ is available in closed form, a high-order quadrature rule such as the Clenshaw--Curtis scheme can be used to evaluate Eq.~\eqref{eq: chf MtM} fast and accurately. However, this direct quadrature-based evaluation of the MtM characteristic function becomes infeasible when the joint density is unavailable or when the dimension of the risk-factor space is large.

To address these limitations,  we employ the COS–CPD expansion introduced in Section~\ref{subsec: COS-CPD} to derive an efficient approximation of $\varphi^V(\cdot)$.

For the portfolio defined in Section \ref{subsec: portfolio setup}, we group the MtM pricing functions into legs sharing common risk factor(s) and thus common state variable(s), and replace the concerned joint density function by its COS-CPD expansion in Eq.~\eqref{eq:CPD form PDF} to obtain
\begin{align}\label{eq: chf MtM CPD}
	\varphi^V(\omega) &\approx\varphi_{\mathrm{cpd}}^V(\omega) \quad \text{with}\nonumber\\
    \varphi_{\mathrm{cpd}}^V(\omega) &:= \int_{\mathbb{R}^{2N + 1}} e^{i\omega  \left[V_0(\gamma_0) + \sum_{n = 1}^{N} V_n(\gamma_n, X_n)\right]} \nonumber\\
    &\qquad \qquad \left[\sum_{r = 1}^{\tilde{R}}  \left(\mathbf{v}_{\gamma_0}^T \A_{\gamma_0}[:,r] \cdot \prod_{n=1}^{N} \mathbf{v}_{\gamma_n}^T \A_{\gamma_n}[:,r] \cdot 
    \mathbf{v}_{X_n}^T \A_{X_n}[:,r] \right)
    \right] \nonumber\\
    & \qquad \qquad \qquad \qquad d\gamma_0  d\gamma_1 \dots d\gamma_N dX_1\cdots dX_N.
\end{align}
Exploiting the separability of the integrand, the above formula can be re-written as
\begin{align}\label{eq: chf MtM CPD split}
	{\varphi}_{\mathrm{cpd}}^V(\omega)&=\sum_{r = 1}^{\tilde{R}} \left( \int_{\mathbb{R}} e^{i\omega  V_0(\gamma_0)}\mathbf{v}_{\gamma_0}^T {\A}_{\gamma_0}[:,r] \ d\gamma_0 \right.
	 \nonumber \\
    &  \qquad  \cdot \;\prod_{n=1}^N  \left. \int_{\mathbb{R}^2}e^{i\omega  V_n(\gamma_n, X_n)} \mathbf{v}_{\gamma_n}^T \A_{\gamma_n}[:,r] \cdot  \mathbf{v}_{X_n}^T \A_{X_n}[:,r] \ d\gamma_n dX_n \right),
\end{align}
where $\A$'s are the factor matrices in the CPD-COS expansion of the joint density function, as in Eq.~\eqref{eq:CPD form PDF}, and \[
\mathbf{v}_{\gamma_n}
:=
\left[
\cos\!\bigl(k \pi \tfrac{\gamma_n-l_{\gamma_n}}{u_{\gamma_n}-l_{\gamma_n}}\bigr)
\right]_{k \in \mathcal{K}_{\gamma_n}}
\in \mathbb{R}^{|\mathcal{K}_{\gamma_n}|\times 1};
\quad
\mathbf{v}_{X_n}
:=
\left[
\cos\!\bigl(k \pi \tfrac{X_n-l_{X_n}}{u_{X_n}-l_{X_n}}\bigr)
\right]_{k \in \mathcal{K}_{X_n}}
\in \mathbb{R}^{|\mathcal{K}_{X_n}|\times 1}.
\]
Thus, the original high-dimensional integral now reduces to a sum of products of one- and two-dimensional integrals, each of which can be evaluated efficiently using a high-order quadrature method, such as Clenshaw--Curtis. Before we present the discretized formula in below, we need to note that the factor matrices are approximations $\tilde{\mathbf{A}}$'s from the offline training stage, which introduces some minor and controlled error, i.e.,
\begin{equation}
\label{eq: tilde chf MtM CPD split}
    \tilde{\varphi}_{\mathrm{cpd}}^V(\omega) \text{ is defined as in Eq.~\eqref{eq: chf MtM CPD split} but with }\mathbf{A} \text{'s replaced by }\tilde{\mathbf{A}}\text{'s}.
\end{equation}

\subsubsection*{The discretized formula}

Let
\[
\{(\gamma_0^{(m)}, w^{\gamma_0}_m)\}_{m=1}^{M_0}
\quad\text{and}\quad
\{((\gamma_n^{(j)},X_n^{(\ell)}), w^{\gamma_n}_j w^{X_n}_\ell)\}_{j=1,\ell=1}^{J_n,L_n}
\]
denote quadrature nodes and weights on $\mathbb{R}$ and $\mathbb{R}^2$,
respectively. The matrices
\[
\mathbf{V}_{\gamma_0} \in \mathbb{R}^{|\mathcal{K}_{\gamma_0}|\times M_0},
\qquad
\mathbf{V}_{\gamma_n} \in \mathbb{R}^{|\mathcal{K}_{\gamma_n}|\times J_n},
\qquad
\mathbf{V}_{X_n} \in \mathbb{R}^{|\mathcal{K}_{X_n}|\times L_n}
\]
collect the COS basis functions evaluated at the corresponding quadrature nodes,
with entries
\[
\mathbf{V}_{\gamma_0}[k,m]
=
\cos\!\Bigl(
k\pi \tfrac{\gamma_0^{(m)}-a_{\gamma_0}}{b_{\gamma_0}-a_{\gamma_0}}
\Bigr),
\]
\[
\mathbf{V}_{\gamma_n}[k,j]
=
\cos\!\Bigl(
k\pi \tfrac{\gamma_n^{(j)}-a_{\gamma_n}}{b_{\gamma_n}-a_{\gamma_n}}
\Bigr),
\qquad
\mathbf{V}_{X_n}[k,\ell]
=
\cos\!\Bigl(
k\pi \tfrac{X_n^{(\ell)}-a_{X_n}}{b_{X_n}-a_{X_n}}
\Bigr).
\]

Replacing the one- and two-dimensional integrals in
Eq.~\eqref{eq: chf MtM CPD split} by quadrature-based approximations, we obtain
\begin{align}
\tilde{\varphi}_{\mathrm{cpd}}^V(\omega)
&\approx \tilde{\varphi}_{\mathrm{cpd,q}}^V(\omega) \quad \mathrm{with}\nonumber \\
\tilde{\varphi}_{\mathrm{cpd,q}}^V(\omega) &:=
\sum_{r=1}^{\tilde R}
\Biggl(
\sum_{m=1}^{M_0}
w^{\gamma_0}_m\,
e^{i\omega V_0(\gamma_0^{(m)})}
\,
\mathbf{V}_{\gamma_0}[:,m]^\top
\tilde{\mathbf{A}}_{\gamma_0}[:,r]
\Biggr)
\nonumber\\
&\qquad\cdot
\prod_{n=1}^N
\Biggl(
\sum_{j=1}^{J_n}\sum_{\ell=1}^{L_n}
w^{\gamma_n}_j w^{X_n}_\ell\,
e^{i\omega V_n(\gamma_n^{(j)},X_n^{(\ell)})}
\,
\bigl(
\mathbf{V}_{\gamma_n}[:,j]^\top
\tilde{\mathbf{A}}_{\gamma_n}[:,r]
\bigr)
\bigl(
\mathbf{V}_{X_n}[:,\ell]^\top
\tilde{\mathbf{A}}_{X_n}[:,r]
\bigr)
\Biggr).
\label{eq: chf MtM CPD split quad}
\end{align}


\begin{remark}[Computational complexity]
Assume $|\mathcal{K}_{\gamma_n}|=K_{\gamma}$, $|\mathcal{K}_{X_n}|=K_X$,
$J_n=J$, and $L_n=L$ for all $n$, and similarly
$|\mathcal{K}_{\gamma_0}|=K_{\gamma_0}$.
Since the cosine-basis matrices are precomputed,  for a fixed $\omega$ the cost of
evaluating~\eqref{eq: chf MtM CPD split quad} scales as
\[
O\!\left(
\tilde R\Bigl(
K_{\gamma_0}M_0
+
N\,(K_{\gamma}J + K_X L + JL)
\Bigr)
\right),
\]
which scales linearly with the number of currencies $N$, the quadrature sizes
$M_0$, $J$, and $L$, as well as with the tensor rank $\tilde R$.
\end{remark}

\begin{remark}
The MtM valuation functions may also be expressed in terms of underlying
state variables when the risk factors themselves are not modeled
explicitly. In this case, the factor matrices correspond to those appearing
in the COS--CPD expansion of the joint density of the state variables. This
representation is adopted in the numerical experiments.
\end{remark}

\begin{remark}
In the offline training stage, the factor matrices need to be
recomputed only upon recalibration of the underlying risk-factor models,
since they encode the joint distribution of the state variables and are
independent of the portfolio composition in terms of financial product
types. In practice, such recalibration typically occurs on a quarterly or
less frequent basis.
\end{remark}

\subsubsection*{Truncation ranges for integrals in Eq.~(\ref{eq: chf MtM CPD split})}
The quadrature rules used in this work require finite integration domains. Since the state-variable distributions typically have unbounded support, we truncate each integration range to control the associated truncation error.

For a given state variable with CDF $F$, we fix a tolerance level $\mathrm{TOL}$ and define the truncation interval as
\begin{equation}\label{eq:truncation_range_integration}
[l,u] := \bigl[F^{-1}(\mathrm{TOL}),\,F^{-1}(1-\mathrm{TOL})\bigr],
\end{equation}
where $F^{-1}$ denotes the percent point function (PPF). By construction, this interval captures at least $1-2\,\mathrm{TOL}$ of the total probability mass. In all numerical experiments, $\mathrm{TOL}=10^{-12}$ was sufficient.

If the distribution of a state variable is known in closed form, the truncation bounds $[l,u]$ are obtained directly from its PPF. If only the characteristic function is available, we first recover the CDF $F$ using the one-dimensional COS method described in Section~\ref{subsec: CDF for Netting}, and subsequently compute $F^{-1}$ numerically via root-finding (e.g., bisection).

\subsubsection{The characteristic function of netting-set exposure}
\label{subsec:chf_nettingset_exposure}
Based on the definition of exposure in Eq.~\eqref{eq:E_nettingset}, the characteristic function of netting-set exposure reads
\begin{equation}
\label{eq: chf exposure netting}
\varphi^{E}(\omega)
=
\mathbb{E}\!\left[e^{i\omega E}\right]
=
\int_{\mathbb{R}^{|\mathbf{Y}|}}
\exp\!\left(i\omega \max\!\left(V(\mathbf{y}),0\right)\right)
f(\mathbf{y})\,d\mathbf{y},
\end{equation}
with $\mathbf{Y}$ being a factor of state variables and $V$ the netting-set MtM value.

The flooring operation introduces a kink at $V(\mathbf{y})=0$, rendering the exposure function only piecewise smooth. As a result, the COS-approximation of the exposure density and CDF exhibit oscillatory artifacts analogous to the classical Gibbs phenomenon in the vicinity of this kink.

For netting-set exposure, this Gibbs phenomenon can be avoided by exploiting the fact that exposure is a deterministic non-negative transformation of the MtM value. Specifically, since $E\ge 0$, the CDF of $E$, denoted by $F^{E}$, satisfies
\begin{equation}
\label{eq: transformation exposure}
F^{E}(e)
=
\begin{cases}
0, & e < 0, \\
F^{V}(e), & e \ge 0,
\end{cases}
\end{equation}
where $F^{V}$ denotes the CDF of the MtM value $V$. Consequently, instead of computing the characteristic function $\varphi^{E}$ of the non-smooth exposure directly, we compute $\varphi^{V}$ and apply the transformation \eqref{eq: transformation exposure} after recovering the MtM CDF using the procedure described in Section~\ref{subsec: CDF for Netting}.

\subsubsection{The characteristic function of counterparty exposure}
\label{subsec:chf_cpty_exposure}
The counterparty exposure is defined as the sum of netting-set exposures across all netting sets associated with the counterparty, cf. Eq.~\eqref{eq:E_cpty}. Accordingly, the characteristic function of the counterparty exposure is given by
\begin{equation}
\label{eq: chf exposure cp}
\varphi^{E^{\mathrm{cpy}}}(\omega)
=
\mathbb{E}\!\left[
\exp\!\left(
i\omega
\sum_{j=1}^{J}
\max\!\left(V^{(j)}(\mathbf{Y}^{(j)}),0\right)
\right)
\right],
\end{equation}
where $V^{(j)}(\mathbf{Y}^{(j)})$ denotes the MtM value of netting set $j$, driven by the corresponding state-variable vector $\mathbf{Y}^{(j)}$.

In contrast to the netting-set case, the transformation \eqref{eq: transformation exposure} is no longer applicable at the counterparty level due to the nonlinearity introduced by summing floored MtM values. As a result, the characteristic function $\varphi^{E^{\mathrm{cpy}}}(\omega)$ must be evaluated numerically.

Formally, Eq.~\eqref{eq: chf exposure cp} can be approximated within the COS--CPD framework by starting from Eq.~\eqref{eq: chf MtM CPD}–\eqref{eq: chf MtM CPD split} and replacing the total MtM contribution in the exponential by the sum of floored netting-set MtM values, that is,
\[
V_0(\gamma_0) + \sum_{n=1}^{N} V_n(\gamma_n,X_n)
\;\mapsto\;
\sum_{j=1}^{J} \max\!\left(V^{(j)}(\mathbf{Y}^{(j)}),0\right).
\]
However, this substitution generally destroys the separability structure that underlies the dimension reduction in Eq.~\eqref{eq: chf MtM CPD split}, particularly when individual netting sets depend on more than just a few state variables. Consequently, a direct COS--CPD-based evaluation of $\varphi^{E^{\mathrm{cpy}}}(\omega)$ is, in general, not computationally feasible. 

A possible remedy is to combine the COS--tensor framework with importance sampling. In particular, a COS--tensor approximation of the total MtM density, or that of individual standalone netting-set exposure, can be used to define an auxiliary density for importance sampling in Monte Carlo simulation of counterparty exposure. The resulting COS--tensor-based importance sampling methodology is beyond the scope of the present paper and will be addressed in future work. However, a brief outline of the main idea is provided in Section~\ref{subsec:COSIS}.

\subsection{CDF and PFE}
Once the characteristic functions are obtained, the one-dimensional COS method is applied to recover the CDF and subsequently the PFE as follows.

\subsubsection{At the netting-set level}
\label{subsec: CDF for Netting}
First, we note that the CDF of the netting-set MtM value can be recovered  by integrating the COS-recovered density, i.e.,
\begin{align}\label{eq:CDF hat}
F_t^{V}(e)
&= \mathbb{P}\!\left(V_t \le e\right) \nonumber\\
&\approx\; \hat{F}_t^{V}(e)
:=
\frac{e-a}{2}\,A_0^{V}(t)
+ (b-a)\sum_{k=1}^{K-1}
\frac{A_k^{V}(t)}{k\pi}
\sin\!\left(
k\pi \frac{e-a}{b-a}
\right),
\end{align}
where $A_k$ for $k=0,\cdots,K-1$ is the COS coefficient as defined in Eq.~\eqref{eq:1D_COS_pdf}.

Next, inserting the numerical approximation of the characteristic function as obtained in Section~\ref{subsec: chf calculation} yields
\begin{equation}\label{eq:CDF approx}
\tilde{F}_t^{V}(e)
:=
\frac{e-a}{2}\,\tilde{A}_0^{V}(t)
+ (b-a)\sum_{k=1}^{K-1}
\frac{\tilde{A}_k^{V}(t)}{k\pi}
\sin\!\left(
k\pi \frac{e-a}{b-a}
\right),
\end{equation}
where we use $\tilde{A}$ to highlight that the characteristic function for the one-dimensional COS is obtained numerically instead of known analytically. That is,
\begin{equation}
\tilde{A}_k^V :=
\frac{2}{b-a}
\operatorname{Re}\!\left\{
 \tilde{\varphi}_{\mathrm{cpd,q}}^V\!\left(\frac{k\pi}{b-a}\right)
e^{-i k \pi a/(b-a)}
\right\},
\qquad k = 0,1,\dots,K-1,
\label{eq:tilde_A_V}
\end{equation} 
with $ \tilde{\varphi}_{\mathrm{cpd,q}}^V(\omega)$ given in Eq.~\eqref{eq: chf MtM CPD split quad}.
Subsequently, the transformation in Eq.~\eqref{eq: transformation exposure} is applied to obtain $\tilde{F}^E_t$, namely the approximation of the CDF of the netting-set level exposure.


Once the exposure CDF is available, the PFE can be computed by numerically solving $F_t(e) - \alpha = 0$, using for example the bisection method. Here $\alpha$ is typically set to $97.5\%$ in practice.

\begin{figure}[H]
\centering
\begin{tikzpicture}[
    font=\small,
    input/.style={
        draw,
        trapezium,
        trapezium left angle=70,
        trapezium right angle=110,
        align=center,
        inner sep=2pt,
        minimum width=2cm
    },
    func/.style={
        draw,
        rectangle,
        rounded corners,
        align=center,
        inner sep=6pt,
        minimum width=2.6cm
    },
    arrow/.style={->, thick},
   node distance=0.6cm and 1.3cm
]

\node[input] (inP) {
\textbf{Input 2:} 
Portfolio file \\
that contains\\
contracts definitions \\
and involved risk factors
};

\node[input, left=of inP] (inA) {
\textbf{Input 1:} \\
Factor matrices from \\
offline training \\(Algorithm~\ref{algo:our_training_method})
};

\node[func, below=of inP] (grp) {
\textbf{Step 1: calculate and group MtM values}\\
(using individual pricing functions \\
and aggregating MtM prices for group \\ 
of contracts that share the same set of state variables)
};

\node[input, right=of inP] (inQ) {
\textbf{Input 3:} \\
Quadrature points\\
and weights per\\
state variable
};

\node[func, below=of grp] (phi) {
\textbf{Step 2: Compute the characteristic function of} \\
\textbf{the netting-set MtM value}\\
(based on COS--CPD expansion of the joint density\\
of all state variables): 
$\tilde{\varphi}^V_{\mathrm{cpd,q}}(\omega)$  from
Eq.~\eqref{eq: chf MtM CPD split quad}
};

\draw[arrow] (inP.south) -- (grp.north);
\draw[arrow] (inA.south) -- (phi.west);
\draw[arrow] (grp.south) -- (phi.north);
\draw[arrow] (inQ.south) -- (phi.east);
\draw[arrow] (inQ.south) -- (grp.east);

\node[func, below=of phi] (Ctilde) {
\textbf{Step 3: Compute 1D-COS coefficients of the} \\
\textbf{ density of the netting-set MtM value:}
$\tilde{C}^V_k$ by
Eq.~\eqref{eq:tilde_A_V}
};

\node[func, below=of Ctilde] (FV) {
\textbf{Step 4: Recover the CDF of the netting-set MtM value}\\
\textbf{based on 1D-COS:} 
$\tilde{F}^V$ by
Eq.~\eqref{eq:CDF approx}
};

\node[func, below=of FV] (FE) {
\textbf{Step 5: Transform to the CDF of the netting-set exposure:} \\
$\tilde{F}^E$ by
Eq.~\eqref{eq: transformation exposure}
};

\draw[arrow] (phi) -- (Ctilde);
\draw[arrow] (Ctilde) -- (FV);
\draw[arrow] (FV) -- (FE);

\end{tikzpicture}
\caption{Computation pipeline from offline-trained factor matrices to exposure distribution,  for a fixed time point in the future.}
\label{fig:cpd_cos_pipeline}
\end{figure}

Other key input parameters of the COS--CPD method in step 3 and 4 above are the truncation upper and lower bounds, i.e., 
$[a,b]$. The method for accurately determining them is described in Section~\ref{subsec: Calculation of the truncation range}.

\subsubsection{At the counterparty level}
\label{subsec: CDF for Counterparty}
At the counterparty level, the total exposure is defined as the sum of multiple floored MtM values across different netting sets. This makes the  counterparty exposure $E_t^{\mathrm{cpt}}$ only piecewise continuous with respect to the underlying state variables. This non-smoothness leads to slow convergence when the CDF of $E_t^{\mathrm{cpt}}$ is recovered directly via the COS expansion  Eq.~\eqref{eq:Fourier-cos_pdf}, even when the characteristic function of the counterparty-level exposure can be evaluated numerically. 

To mitigate the Gibbs phenomenon, a simple but effective solution is to employ spectral filters as discussed in \cite{ruijter2015application}. Such filters operate directly in the Fourier domain and are known to improve convergence near discontinuities and non-smoothness points without introducing additional computational overhead.


The filtered COS approximation of the CDF of $E_t^{\mathrm{cpt}}$ has the same form as in~\eqref{eq:CDF approx}, except that each Fourier--cosine coefficient is multiplied by a spectral filter $\sigma(\tfrac{k}{K-1})$. The filter $\sigma(\cdot)$ is a real-valued, even function supported on the interval $[-1,1]$ and is chosen to satisfy suitable convergence and regularity properties; see, for example, \cite{Vandeven1991,gottlieb1997gibbs}. The procedure for determining the truncation interval $[a,b]$ is described in Section~\ref{subsec: Calculation of the truncation range}.

Afterwards, PFE is determined the same way as in the netting-set case.


\subsubsection{The truncation range required for  the CDF calculation}
\label{subsec: Calculation of the truncation range}

The choice of the COS truncation interval $[a,b]$ plays a crucial role in the
accuracy of the recovered CDF. 
%
%
A cumulant-, or central-moment-based rule of thumb for selecting $[a,b]$ was proposed in~\cite{fang2009novel}. For portfolio MtM values or exposures, the central moments are generally not available in closed form. We therefore adopt a simplified version of truncation range definition that involves only the first two central moments and compute them numerically. That is,
\begin{equation}
	\label{eq: truncation range}
	[a,\, b] = \left[c_1 - L\,\sqrt{c_2}, \; c_1 + L\,\sqrt{c_2} \right],
\end{equation}
where $L$ is typically chosen between 8 and 10,  and $c_1$ and $c_2$ denote the first and the second central moment, respectively. For netting-set–level exposure or MtM calculations, the target variable is the netting-set MtM value $V(\mathbf{Y}_t)$, whereas for counterparty-level exposure, the target random variable is $E^{\mathrm{cpt}}_t$; see Section~\ref{subsec:chf_nettingset_exposure}. In all numerical experiments reported in this paper, we set $L=8$.

The central moments are computed using the same procedure employed for evaluating the corresponding characteristic function. Let us again fix a future time point $t$ and suppress the time index. The $k$-th central moment of the portfolio MtM value is given by
\begin{equation}
    \mathbb{E}\!\left[V^k(\mathbf{Y})\right]
    =
    \int_{\mathbb{R}^{|\mathbf{Y}|}}
    V^k(\mathbf{y})\, f(\mathbf{y})\, d\mathbf{y}.
\end{equation}
For $k=1,2$, the separability of state variables in $V^k(\mathbf{Y})$ is preserved. Consequently, the same COS--CPD-based techniques as discussed in Section~\ref{subsec: chf calculation} can be applied to evaluate these moments efficiently.
It is important to note that the computational cost of estimating the first two moments is substantially lower than that of evaluating the full characteristic function. In contrast to the characteristic function, which must be computed with high accuracy over a wide range of frequencies, the moments are only used to determine a rough truncation interval for the COS expansion. As a result, their computation tolerates significantly coarser approximations across all components of the COS–CPD framework, including a small number of cosine terms, a low tensor rank, and a limited number of quadrature points.


For counterparty-level exposure, we set $a=0$, since exposure is floored at zero by definition.
The upper bound still relies on the first two moments, which must be computed via quadrature rules directly due to the inseparability of state variables inside max operator, as discussed in Section~\ref{subsec:chf_cpty_exposure}. This remains computationally inexpensive for a moderate number of state variables.

\subsection{EE and EE sensitivities}\label{subsec:EE and sensitivities}

In this subsection, we demonstrate that EE and EE sensitivities can be computed efficiently within the COS-tensor framework, alongside PFE, once the characteristic function is available. For clarity of presentation, herewith we omit the superscript notation that distinguishes netting sets from counterparty portfolios, since the methodology applies to both cases in the same way.

\subsubsection{The COS formula for EE}
Define $L := \max(a,0)$ and $U := \max(b,0)$, where $[a,b]$ is the truncation range for the portfolio value $V_t$. With this change of variables, \eqref{eq: EE general} becomes
\begin{equation}
	\label{eq: EE portfolio}
    \mathrm{EE}(t) = \mathbb{E}[ \max(V_t,0) | \mathcal{F}_0]
     = \int_{\mathbb{R}} \max(v,0)\, f_t^V(v)\, dv
	\approx \int_L^U v \,f_t^V(v)\, dv,
\end{equation}
where  $f_t^V$ denotes the density function of $V_t$. 
The structure of the above formula closely resembles that of pricing a European call option, with $\max(V_t,0)$ playing the role of payoff. Indeed, EE can be efficiently calculated using the 
COS method, once the characteristic function of the total MtM value of the portfolio is computed, for example using the method developed in Section~\ref{subsec: chf calculation}. Below we present the explicit COS  formula ready for implementation:
\begin{equation}
\label{eq:EE_COS_general}
    \mathrm{EE}(t) \approx \sum_{k=0}^{K-1}{}' A_k(t)\, W_k(L,U),
\end{equation}
with $A_k$ given in~(\ref{eq:approximated COS coef}). The time dependence of $A_k$ is a result of that of the characteristic function. Using the notation $\omega_k=\frac{k\pi}{b-a}$, the weights are given by
\begin{equation}\label{eq:Wk_general}
W_k(L,U) =
\begin{cases}
    \tfrac{1}{2}(U^2 - L^2), 
        & k = 0, \\[1.0ex]
    \left[ \dfrac{v \sin\!\big(\omega_k(v-a)\big)}{\omega_k}
    + \dfrac{\cos\!\big(\omega_k(v-a)\big)}{\omega_k^2} \right]_{v=L}^{U}, 
        & k \geq 1.
\end{cases}
\end{equation}

\subsubsection{Calculation of EE sensitivities}

%
%

EE is a key input for xVA. Sensitivities of xVA with respect to underlying risk factors, often referred to as \emph{xVA deltas}, are essential for risk management, hedging, and capital allocation. EE sensitivities therefore constitute essential inputs for the calculation of xVA deltas.

To illustrate the methodology, we consider the sensitivity of EE w.r.t. the \(n\)-th foreign interest rate. Differentiating the COS expansion of EE in \eqref{eq:EE_COS_general} w.r.t. \(\gamma_n(0)\) yields

\begin{equation}
\label{eq:EE_grad_master}
\frac{\partial \mathrm{EE}(t)}{\partial \gamma_n(0)}
\approx\sum_{k=0}^{K-1}{}' \frac{\partial A_k(t)}{\partial \gamma_n(0)}W_k(L,U),
\;\text{with}\;
\frac{\partial A_k(t)}{\partial \gamma_n(0)}
=\frac{2}{b-a}\operatorname{Re}\!\left\{e^{-i\omega_k a}\,
\frac{\partial \varphi(\omega_k,\gamma_n(0),t)}{\partial \gamma_n(0)}\right\},
\end{equation}
where $\omega_k=\frac{k\pi}{b-a}$ and $W_k$ is given in~(\ref{eq:Wk_general}), and $\gamma_n(0)$ denotes the initial value of the $n-$th foreign interest rate. Sensitivities with respect to other risk factors can be derived analogously. 

Hence, the computation of EE sensitivities reduces to the evaluation of derivatives of the characteristic function w.r.t. the relevant initial values of the risk factors. In the following, we distinguish two cases depending on whether the initial value of a risk factor can be separated from the distributional randomness of the underlying stochastic process.

\paragraph{Change-of-variables approach.}
For many stochastic process models, the dependence on the initial state can be isolated through an appropriate change of variables. In particular, in view of the portfolio decomposition in~(\ref{eq:portfolio_split}), the dependence of the portfolio value \(V\)—and hence of the exposure \(\mathrm{EE}\)—on the initial value of a given risk factor, for example \(\gamma_n(0)\), arises solely through the corresponding portfolio component \(V_n\). Consequently,
\[
\frac{\partial V}{\partial \gamma_n(0)} = \frac{\partial V_n}{\partial \gamma_n(0)} .
\]

Moreover, the component \(V_n\) can be reparameterized so as to admit a representation of the form
\[
    V_n\big(\gamma_n(t), X_n(t), t\big)
    = V_n\big(g(\gamma_n(0), z), X_n(t), t\big), 
    \qquad z \sim f_Z,
\]
where \(z\) is an auxiliary random variable, independent of the initial level \(\gamma_n(0)\), with distribution \(f_Z\). Such representations naturally arise, for example, for short rate in zero-coupon bond pricing—and hence for the MtM of linear interest-rate products—under general affine term structure models.

Under this transformation, differentiating the characteristic function of \(V\) with respect to \(\gamma_n(0)\) yields
\begin{equation}\label{eq:phi_deriV_0eparam}
    \frac{\partial \varphi_V}{\partial \gamma_n(0)}(\omega)
    =
    i\omega\,\mathbb E_Z\!\left[
        \exp\!\Big(i\omega \sum_{i=0}^N V_i\Big)\,
        \frac{\partial V_n\!\big(g(\gamma_n(0),z),t\big)}{\partial g}\,
        \frac{\partial g(\gamma_n(0),z)}{\partial \gamma_n(0)}
    \right].
\end{equation}

Sensitivities with respect to the initial values of the FX rates can be derived analogously.

Note that the separability of \(V\) is preserved in the calculation, which enables the treatment of sensitivities in high dimensional settings using the tensor decomposition techniques described in Subsection~\ref{subsec: chf calculation}.

\paragraph{Score-function approach.}
If such a separation of the initial value is not available, we exploit the fact that \(\gamma_n(0)\) affects the characteristic function of \(V\) only through the joint density of the risk factors, and not through the pricing function itself. In this case, the derivative of the characteristic function can be written as
\begin{equation}\label{eq:phi_deriv_score}
    \frac{\partial \varphi_V}{\partial \gamma_n(0)}(\omega)
    =
    \int_{\mathbb R^{|\mathbf Y_t|}} 
    e^{i\omega V(\mathbf y_t,t)}\,
    \frac{\partial f(\mathbf y_t; \gamma_n(0))}{\partial \gamma_n(0)}\,
    d\mathbf y_t
    =
    \mathbb E\!\left[
        e^{i\omega V(\mathbf Y_t,t)}\,
        s(\mathbf Y_t; \gamma_n(0))
    \right],
\end{equation}
where 
\( s(\mathbf y; \gamma_n(0)) = \partial_{\gamma_n(0)} \log f(\mathbf y; \gamma_n(0)) \)
denotes the score function of the joint density.

The gradient term \( \partial_{\gamma_n(0)} f(\mathbf Y_t; \gamma_n(0)) \) can either be evaluated analytically, when available, or approximated numerically by differentiating the truncated multi-dimensional Fourier--cosine representation of $f$ introduced in Section~\ref{subsec: COS method}, whose coefficient tensor is represented in low rank (e.g., via CPD). In both cases, a separable low-rank approximation of the gradient can be constructed and substituted into \eqref{eq:phi_deriv_score}, enabling an efficient evaluation of the resulting integral.

This score-function approach is more general than the change-of-variables method, but is typically less efficient in practical implementations.

\begin{remark}
The above derivations are stated at the netting-set level. At the counterparty level, EE sensitivities are obtained by summing the sensitivities of all constituent netting sets, since counterparty exposure is defined as their sum.
\end{remark}

\section{Numerical tests}
\label{sec: numerical_tests}
This section presents the numerical experiments conducted to assess the accuracy and computational efficiency of the proposed COS--tensor framework for PFE estimation. 
We benchmark the performance against MC simulation method. The tests span both netting-set and counterparty-level exposures, as well as low- and high-dimensional model configurations.

\subsection{Model setup}\label{subsec: Model setup}

All experiments are conducted under a hybrid Hull--White and geometric Brownian motion (HW--GBM) framework, which is widely used in industry for modeling interest rate (IR) and foreign exchange (FX) risk factors. This hybrid model combines mean-reverting short rate dynamics with lognormally distributed FX rates. Our implementation closely follows the specification in \cite{piterbarg2005multi}, with tractable simplifications drawn from \cite{di2012general}. The setup provides a flexible yet analytically manageable platform for evaluating CCR exposure.

The short rates are modeled as Ornstein--Uhlenbeck processes with mean-reversion rate $a_i$ and volatility parameter $\sigma_i$:
\begin{gather*}
    a_{\gamma_0} = 1\%, \quad a_{\gamma_1} = 5\%, \quad a_{\gamma_2} = 2\%, \quad a_{\gamma_3} = 1\%, \\
    \sigma_{\gamma_0} = 0.7\%, \quad \sigma_{\gamma_1} = 1.2\%, \quad \sigma_{\gamma_2} = 0.6\%, \quad \sigma_{\gamma_3} = 0.8\%.
\end{gather*}
Here, ${\gamma_0}$ denotes the domestic currency (USD), and $\gamma_1$, $\gamma_2$, and $\gamma_3$ correspond to JPY, EUR, and GBP, respectively.

The corresponding market discount functions are assumed to follow deterministic exponential curves:
\begin{equation*}
\begin{split}
P^M_{\gamma_0}(0,T) = \exp(-0.02T), \quad P^M_{\gamma_1}(0,T) = \exp(-0.05T), \\
P^M_{\gamma_2}(0,T) = \exp(-0.05T), \quad P^M_{\gamma_3}(0,T) = \exp(-0.04T).
\end{split}
\end{equation*}

Each exchange rate follows a standard GBM process with drift $\mu_i$ and volatility $\sigma_i$:
\begin{gather*}
    \mu_{X_1} = 0.8\%, \quad \mu_{X_2} = 0.6\%, \quad \mu_{X_3} = 1\%, \\
    \sigma_{X_1} = 2\%, \quad \sigma_{X_2} = 1.5\%, \quad \sigma_{X_3} = 3\%.
\end{gather*}
We denote the USD/JPY, USD/EUR, and USD/GBP exchange rates as $X_1$, $X_2$, and $X_3$, respectively.

Initial spot FX rates are set to:
\[
X_1 = 105, \quad X_2 = \frac{1}{1.35}, \quad X_3 = 0.7732,
\]
where the first two are aligned with the settings in \cite{grzelak2012cross}, and the last reflects the market value on April 24, 2025.

The risk factors are coupled via a static correlation matrix reflecting stylized dependencies commonly used in risk modeling. The interest rates are moderately correlated, and all FX rates share positive correlation among themselves but are negatively correlated with the interest rate processes:
\begin{equation*}
	\begin{aligned}
		\rho_{\gamma_i, \gamma_j} = 0.25, \quad \text{for all } i \ne j, \\
		\rho_{X_i, X_j} = 0.60, \quad \text{for all } i \ne j, \\
		\rho_{\gamma_i, X_j} = -0.15, \quad \text{for all } i, j.
	\end{aligned}
\end{equation*}

Testing portfolios are generated using a Python script, which randomly creates linear derivatives (e.g., IR forward rate agreement, IR swap, FX forward, and cross currency swap) with a random currency, notional, fixed rate, tenor, and maturity.

For all the experiments, CPU times are recorded on a single threaded implementation using an Intel Core i7-10700F processor at 2.90GHz and a 2048 KB L1 cache. Timing is reported in seconds.

\subsection{Training performance}\label{subsec: training performance}
Efficient training of low-rank tensor approximations is critical in high-dimensional risk modeling, where scalability and accuracy are often in tension. To evaluate the effectiveness of our proposed Fourier-domain training method, we benchmark it against the physical-domain approach reviewed in Section~\ref{subsec:physical_training}.

In the physical-domain setting, the CPD factor matrices are trained by minimizing the sample mean-squared error between the target function values $f(\mathbf{x}^m)$ and the corresponding CPD-based Fourier--cosine approximation evaluated at the same inputs, cf.\ Eq.~\eqref{eq: optimization real}. In each ALS step, we implement the resulting stochastic optimization using the ADAM optimizer \cite{kingma2014adam}, a variant of SGD that incorporates momentum and adaptive learning rates based on running estimates of the first and second moments of the gradient. While the original ADAM employs a step-size scaling proportional to the inverse square root of the iteration count, we adopt an exponentially decaying step size, which empirically yields faster convergence in our setting.

The physical-domain training inputs are constructed as a Cartesian product of equidistant grids. In a $d$-dimensional setting with $K$ retained cosine terms per dimension, we use $m=K$ grid points per dimension, resulting in $M=K^d$ evaluation points $\{\mathbf{x}^m\}_{m=1}^{K^d}$. This choice balances model flexibility and data constraints: $K$ determines the number of cosine coefficients to be approximated per dimension, while $m$ controls the number of pointwise constraints in the supervised objective. Empirical analysis indicates that $m<K$ leads to an under-constrained fit and $m>K$ to over-parameterized behaviour with deteriorating global approximation quality, whereas the balanced choice $m=K$ yields stable and accurate recovery on the truncation domain.

In contrast, the Fourier-domain training method operates directly on sampled fibers of the cosine coefficient tensor $\mathcal{A}_{\mathbf{k}}$, evaluated on-the-fly from the characteristic function $\varphi$. The same ALS framework is used, but each sampled least-squares subproblem \eqref{eq:als_subproblem} is solved by a randomized least-squares procedure based on QR decomposition. No density evaluations are required.

To assess performance, we evaluate convergence in both iteration count and wall-clock time. Convergence is measured using the infinity norm of the coefficient error tensor $\epsilon := \mathcal{A}_{\mathbf{k}} - \widetilde{\mathcal{A}}_{\mathbf{k}}$, where $\widetilde{\mathcal{A}}_{\mathbf{k}}$ denotes the rank-$\tilde R$ CPD reconstruction implied by the trained factor matrices. Since forming the full tensor is infeasible in high dimensions, we restrict this test to a three-dimensional setting involving USD, JPY, and the USD/JPY exchange rate.

\begin{figure}
    \centering
    \includegraphics[width=0.7\linewidth]{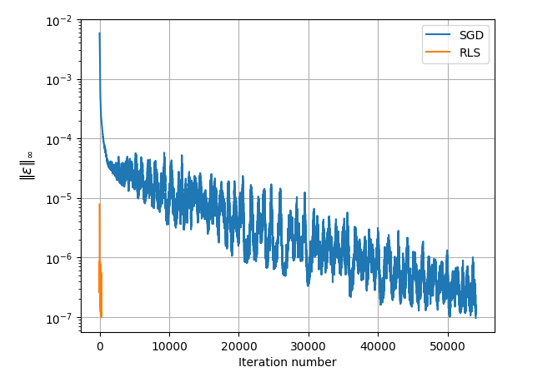}
    \caption{Convergence behavior of physical-domain ADAM-based training and Fourier-domain RLS-based training in terms of the absolute coefficient error measured by the infinity norm, $\|\epsilon\|_\infty$, over iterations for a 3-dimensional setting.}
    \label{fig: training iterations}
\end{figure}

For both methods, training is conducted using mini-batches of size $M=900$. The target cosine tensor is approximated using a rank-$\tilde R$ CP decomposition with $\tilde R=15$ and $K=32$ cosine terms per dimension. Figure~\ref{fig: training iterations} illustrates the convergence behavior per iteration. The Fourier-domain training converges within a few hundred iterations, whereas the physical-domain training requires substantially more iterations to reach comparable error levels.

Iteration count alone does not reflect computational efficiency. In the Fourier-domain method, each ALS update involves solving an overdetermined least-squares problem based on dynamically sampled tensor entries, whereas in the physical-domain benchmark the updates are gradient-based and pointwise. Table~\ref{tab:comparison cpu training} reports the total CPU time required to achieve prescribed error thresholds. All timings were obtained on the hardware described in Section~\ref{subsec: Model setup}.

\begin{table}[h]
\centering
\caption{CPU time (in seconds) required for physical-domain ADAM-based training and Fourier-domain RLS-based training to achieve $\|\epsilon\|_\infty < 10^{-x}$ for various thresholds in a 3-dimensional setting.}
\label{tab:comparison cpu training}
\begin{tabular}{lrrr}
\hline
\textbf{Method} & \begin{tabular}[c]{@{}c@{}}\\ $\|\epsilon\|_\infty < 10^{-5}$\end{tabular} 
       & \begin{tabular}[c]{@{}c@{}}\textbf{CPU Time (s)}\\ $\|\epsilon\|_\infty < 10^{-6}$\end{tabular} 
       & \begin{tabular}[c]{@{}c@{}}\\ $\|\epsilon\|_\infty < 10^{-7}$\end{tabular} \\ \hline

\textbf{ADAM}    & 98.3           & 365.2          & 1062.0         \\
\textbf{RLS}    & 0.04           & 0.1            & 11.9           \\
\hline
\end{tabular}
\end{table}
The results show that Fourier-domain training achieves similar accuracy levels with significantly lower computational cost, with speedups of up to two orders of magnitude relative to physical-domain ADAM-based training. These gains reflect the structural advantages of exploiting the characteristic function to access cosine coefficients directly, thereby avoiding repeated density evaluations and enabling efficient low-rank training in the Fourier domain.

\subsection{PFE at the netting--set level}\label{subsec: PFE netting}
We first assess the convergence behavior and computational efficiency of the COS method for computing PFE at the netting–set level. In this experiment, we consider a three‑dimensional setting comprising the USD and JPY interest‑rate factors and the USD/JPY exchange rate. Reference values are computed using the COS method with conservative numerical settings: $150$ Fourier expansion terms, $130$ quadrature points per state variable, and a truncation tolerance $\mathrm{TOL} = 10^{-12}$, as defined in Section~\ref{subsec: Calculation of the truncation range}. The convergence tests use $t = 8.6$ years, corresponding to a third of the maximum maturity of the portfolio.

\begin{figure}
	\centering
	\begin{subfigure}[b]{0.49\textwidth}
		\centering
		\includegraphics[width=\textwidth]{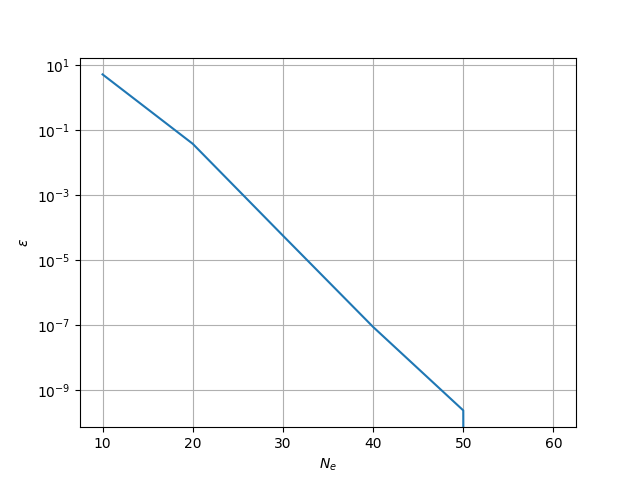}
		\caption{Error vs. number of expansion terms}
		\label{fig: convergence netting expansion}
	\end{subfigure}
	\hfill
	\begin{subfigure}[b]{0.49\textwidth}
		\centering
		\includegraphics[width=\textwidth]{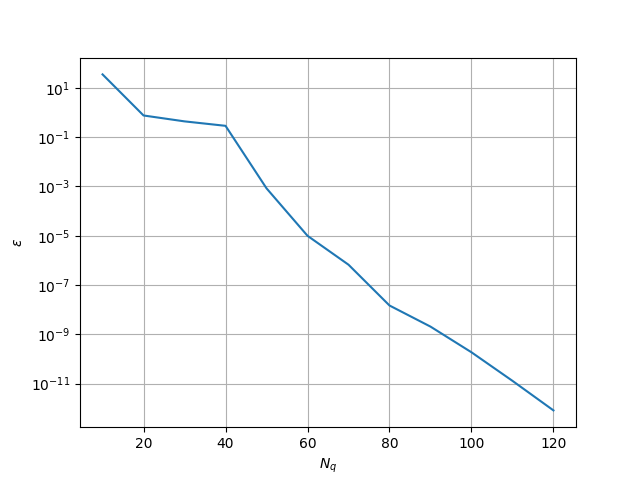}
		\caption{Error vs. number of quadrature points}
		\label{fig: convergence netting quadrature}
	\end{subfigure}
    \caption{Convergence of PFE estimates at the netting–set level for a 3-dimensional portfolio of 1000 linear IR and FX derivatives. The relative error, expressed as a percentage of the benchmark PFE value $\$11445.06$, is plotted on a logarithmic scale.}
	\label{fig: convergence netting}
\end{figure}

Figure~\ref{fig: convergence netting} presents convergence with respect to (a) the number of expansion terms, and (b) the number of quadrature points. Both present exponential convergence, with machine precision attained using $60$ expansion terms.

To benchmark the COS method against MC simulation, we compute PFE at two exposure dates ($T_{\max}/3$ and $2T_{\max}/3$) for portfolios with $1,000$ and $10,000$ derivatives. The MC path counts are chosen to reflect practical use cases, where relative errors under $1\%$ are generally sufficient. For the COS method, we use a fixed configuration of $32$ expansion terms and $50$ quadrature points per state variable. These settings are reused in the counterparty-level experiments to ensure comparability and to isolate methodological differences from parameter tuning effects.

\begin{table}[h!]
\caption{Accuracy and computational time required to calculate the PFE of 3-dimensional netting sets with $1000$ and $10,000$ derivatives. Relative errors (in $\%$) are averaged over two exposure dates.}
\label{table:comptime_netting_combined}
\centering
\begin{tabular}{@{} l r S[table-format=1.2e-1, round-mode=places, round-precision=2] @{}}
\toprule
Method & {CPU Time (s)} & {Error (\%)} \\
\midrule
\multicolumn{3}{l}{\emph{Portfolio with 1,000 derivatives}} \\
MC ($0.5 \cdot 10^5$) & 1.5  & 0.315 \\
MC ($10^5$)            & 3.1  & 0.245 \\
MC ($5 \cdot 10^5$)    & 16.7 & 0.094 \\
COS                    & 0.6  & 8.734e-5 \\ 
\midrule
\multicolumn{3}{l}{\emph{Portfolio with 10,000 derivatives}} \\
MC ($0.5 \cdot 10^5$) & 13.3  & 1.373 \\
MC ($10^5$)            & 26.4  & 0.989 \\
MC ($5 \cdot 10^5$)    & 147.2 & 0.399 \\
COS                    & 1.2   & 6.681e-4 \\
\bottomrule
\end{tabular}
\end{table}
These results highlight the superior computational efficiency of the COS method. MC runtimes grow nearly linearly with both path count and portfolio size, while COS remains highly efficient due to its low number of required function evaluations. For a portfolio of $10,000$ derivatives, the COS method takes just $1.2$ seconds, compared to the $147.2$ seconds with MC at lower accuracy. This corresponds to a speedup of more than two orders of magnitude. Similar gains are observed in the $1,000$-derivative case, where COS is roughly $25$ times faster. For both methods, the increase in error with portfolio size is modest and expected for the accumulation of small local errors across many contracts.

\subsection{Test results for high-dimensional netting sets}
In Subsection~\ref{subsec: PFE netting}, we analyzed the convergence of the COS method with respect to the number of quadrature points used for ch.f. evaluation and the number of expansion terms used to approximate the CDF. In higher-dimensional settings, two additional sources of approximation error arise: the tensor rank and the number of cosine expansion terms used in the Fourier representation of the joint PDF, as discussed in Section~\ref{subsec: chf calculation}.

\begin{figure}[h!]
	\centering
	\begin{subfigure}[b]{0.49\textwidth}
		\centering
		\includegraphics[width=\textwidth]{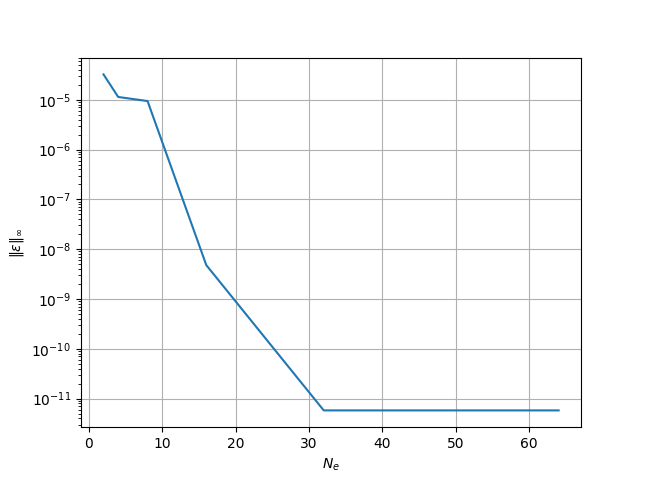}
		\caption{Uncorrelated case ($R=1$): empirical density error vs.\ number of cosine expansion terms $N_e$.}
		\label{fig: convergence expansion pdf}
	\end{subfigure}
	\hfill
	\begin{subfigure}[b]{0.49\textwidth}
		\centering
		\includegraphics[width=\textwidth]{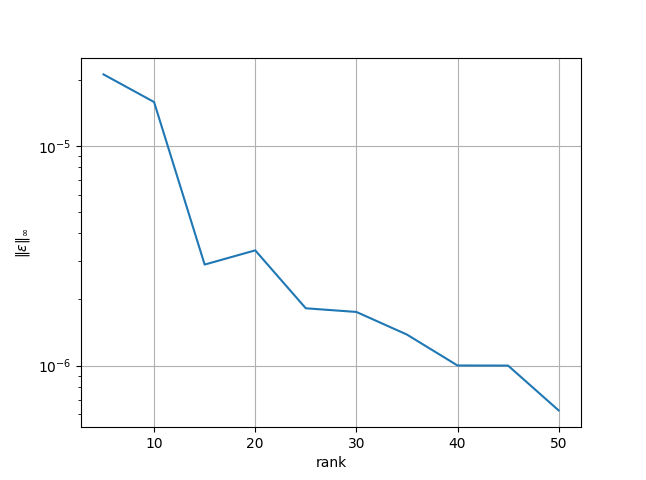}
		\caption{Correlated case: empirical density error vs.\ CP rank $R$.}
		\label{fig: convergence rank}
	\end{subfigure}
	\caption{Convergence of the COS--CPD approximation for recovering a seven-dimensional Gaussian density. The benchmark is the analytic multivariate normal density evaluated on a fixed test set of $M=5000$ points on the truncation domain $D=[a,b]^7$, constructed as in~\eqref{eq:truncation_range_integration} with $TOL=10^{-8}$. The vertical axis reports the empirical $\ell^\infty$ error and is displayed on a logarithmic scale.}
	\label{fig: convergence with CPD}
\end{figure}

To isolate the effect of the cosine truncation in the tensor representation, we first consider an uncorrelated seven-dimensional standard normal model. In this case the joint density factorises and admits an exact rank-$R=1$ CP representation. We draw a fixed randomized Sobol test set of $M=5000$ points on the hyperrectangular truncation domain $D=[a,b]^7$, where $[a,b]$ is constructed as in~\eqref{eq:truncation_range_integration} with $TOL=10^{-8}$. For each truncation level $N_e$, we evaluate the COS--CPD approximation $\hat f_{\mathrm{CPD}}$ and compare it to the analytic Gaussian benchmark $f$ on the same test set. Figure~\ref{fig: convergence expansion pdf} reports the empirical $\ell^\infty$ error.

Next, we reintroduce correlation among the seven risk factors and investigate the sensitivity of the COS--CPD approximation to the CP rank. Using the same test set construction on $D$, we compute the correlated Gaussian benchmark density $f$ and the rank-$R$ COS--CPD approximation $\hat f_{\mathrm{CPD}}$. In this experiment, the number of cosine expansion terms is fixed at $N_e=32$, so that the effect of tensor rank can be isolated. We report the same empirical $\ell^\infty$ error as a function of $R$; see Figure~\ref{fig: convergence rank}.

Figure~\ref{fig: convergence expansion pdf} shows exponential decay of the empirical error as $N_e$ increases, consistent with the smoothness of the Gaussian density and the spectral convergence properties of cosine expansions on a bounded interval. Figure~\ref{fig: convergence rank} exhibits non-monotone behaviour in $R$. While the best rank-$R$ approximation error is non-increasing in $R$ by nestedness of the model class, the CP fitting problem is non-convex and can be ill-conditioned. Consequently, a practical optimization routine may converge to suboptimal stationary points or suffer from numerical instability, leading to non-monotone error curves across ranks.

We now assess COS performance in a seven-dimensional setting involving four currencies and three exchange rates, as defined in the model setup. As a visual check on the accuracy of the COS method in a realistic high-dimensional setting, we compare the full PFE profile obtained from COS to that from MC simulation for the $10{,}000$-derivative portfolio with $7$ risk factors. The MC profile, computed using $50{,}000$ simulation paths, is displayed at the $97.5\%$ quantile together with $95\%$ pointwise order-statistic confidence bands, while the COS estimates are evaluated at the same $50$ equally spaced time points.

\begin{figure}[h!]
    \centering
    \includegraphics[width=0.7\linewidth]{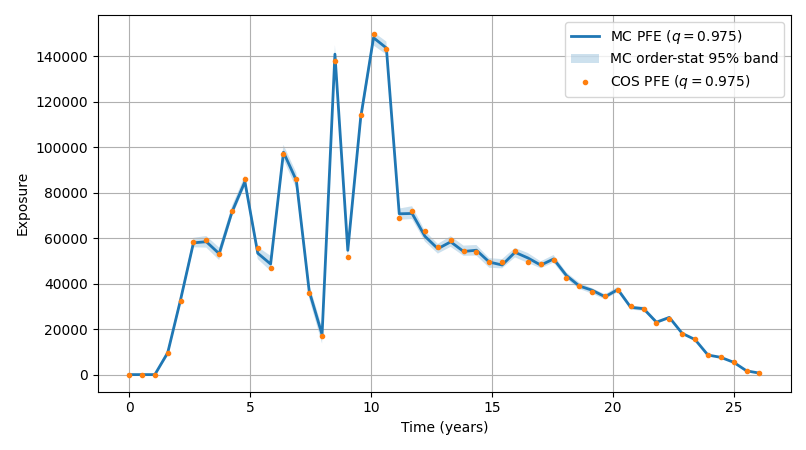}
    \caption{Potential Future Exposure (PFE) profile for a $10{,}000$-derivative portfolio with $7$ risk factors, evaluated over $50$ time points. The solid blue line shows the Monte Carlo (MC) PFE at the $97.5\%$ quantile, computed using $50{,}000$ simulation paths, with shaded $95\%$ pointwise order-statistic confidence bands. Orange dots represent the COS method PFE at the same quantile. }
    \label{fig:pfe_profile_10kder}
\end{figure}

Figure~\ref{fig:pfe_profile_10kder} shows that the COS profile closely tracks the MC profile at all maturities, with all points lying within the MC confidence bands. This visual agreement provides additional evidence that the COS method produces results consistent with the MC approach.

We further evaluate the accuracy of the COS method for two portfolios containing $1{,}000$ and $10{,}000$ derivatives. For each portfolio, the PFE is computed at two time points: one-third and two-thirds of the maturity of the longest contract. Due to the high dimensionality, direct COS-based reference values are infeasible. Instead, benchmark values are computed using MC simulation with $10^7$ scenarios. The COS method employs $32$ expansion terms for the CDF, $50$ quadrature points per dimension, $32$ expansion terms in the joint PDF approximation, and CP rank $R = 30$.

\begin{table}[h!]
\caption{Accuracy and computational time required to calculate the PFE of 7-dimensional netting sets with $1000$ and $10,000$ derivatives. Relative errors (in $\%$) are averaged over two exposure dates.}
\label{table:comptime_highdim_combined}
\centering
\begin{tabular}{@{} l r S[table-format=1.2e-1, round-mode=places, round-precision=2] @{}}
\toprule
Method & {CPU Time (s)} & {Error (\%)} \\
\midrule
\multicolumn{3}{l}{\emph{Portfolio with 1,000 derivatives}} \\
		MC $(0.5 \cdot 10^5)$ & 1.4            & 1.938              \\
		MC $(10^5)$           & 2.8           & 1.103             \\
		MC $(5 \cdot 10^5)$   & 15.5            & 0.560            \\
		COS                 & 1.0             & 0.876             \\
\midrule
\multicolumn{3}{l}{\emph{Portfolio with 10,000 derivatives}} \\
		MC $(0.5 \cdot 10^5)$ & 15.1             & 0.851            \\
		MC $(10^5)$           & 30.5            & 0.721            \\
		MC $(5 \cdot 10^5)$   & 170.2          & 0.408           \\
		COS                 & 1.7             & 0.122              \\
\bottomrule
\end{tabular}
\end{table}

Table~\ref{table:comptime_highdim_combined} confirms the expected linear growth of MC runtimes with portfolio size. Increasing the number of derivatives by a factor of ten results in a corresponding ten-fold increase in CPU time. In contrast, the COS-CPD  method exhibits only a modest runtime increase, from $1.0$ to $1.7$ seconds, demonstrating scalability in high-dimensional settings.

Accuracy remains stable across both portfolio sizes. In fact, the larger portfolio exhibits slightly lower errors, which is natural since adding more derivatives smooths the aggregated exposure distribution through diversification. Notably, for the $10{,}000$-derivative portfolio the COS method achieves an error of $0.12\%$ while requiring only a fraction of the time needed for MC simulation. These results underscore the advantages of the COS–tensor framework: it offers significant computational savings without sacrificing accuracy, even as the dimension and size of the portfolio increase.

\subsection{Sensitivities}
We compute netting-set EE sensitivities for a three-dimensional problem setup consisting of a portfolio of $1{,}000$ derivatives at two exposure dates. Reference values are obtained with a high-accuracy COS configuration (150 expansion terms; 130 quadrature points per state variable), following Section~\ref{subsec:EE and sensitivities}. 

For MC, we do not employ shock-and-revalue. Instead, we use the same derivative principle as in the COS approach. Let $\theta$ denote any of the initial values of the risk factors (e.g.\ $\gamma_0, \gamma_1,$ or $X_1$). The pathwise sensitivity is then estimated as
\begin{equation}
\label{eq:mc_pathwise_ee_greek}
\widehat{\partial_\theta \mathrm{EE}}(t)
=\frac{1}{N}\sum_{p=1}^{N}\mathds{1}_{V_t(\mathbf{Y}_t^{(p)})>0}\,
\partial_\theta V_t(\mathbf{Y}_t^{(p)}),
\end{equation}
where $\partial_\theta V_t$ is obtained analytically per path using the closed-form state derivatives (and chain rule) described in Section~\ref{subsec:EE and sensitivities}. 
This eliminates finite differences and avoids additional revaluations.

Errors are reported as differences from the reference values. The COS method uses 32 expansion terms and 50 quadrature points per state variable. A comprehensive presentation of all CPU times and accuracy results is provided in Table \ref{table: comptime EE MC COS netting}.

\begin{table}[h!]
	\caption{Accuracy and computational time required to calculate EE sensitivities of a 3-dimensional netting set with $1{,}000$ derivatives. Relative errors (in \%) are averaged over two exposures dates.}
	\label{table: comptime EE MC COS netting}
	\centering
	\begin{tabular}{lrrrr}
		\hline
		&        & Error (\%)        &        &                      \\
		Method  & CPU Time (seconds) &  $\frac{\partial \mathrm{EE}}{\partial \gamma_0}$  & $\frac{\partial \mathrm{EE}}{\partial \gamma_1}$ & $\frac{\partial \mathrm{EE}}{\partial X_{1}}$ \\ \hline
		MC ($0.5 \cdot 10^5$) 	& 2.4 & 0.1681 & 0.1341  & 0.9949  \\
		MC $(10^5)$ 			& 5.0 & 0.1258  & 0.0786  & 0.5421  \\
		MC $(5 \cdot 10^5)$ 	& 27.9 & 0.0434  & 0.0424  & 0.5046  \\
		COS 					& 0.8  & $3 \cdot 10^{-5}$  & $5 \cdot 10^{-5}$ &  $9 \cdot 10^{-5}$ \\ \hline
	\end{tabular}
\end{table}

Table \ref{table: comptime EE MC COS netting} reports a single CPU time for all sensitivities due to the equal computational complexity involved in their calculations. Furthermore, the findings reaffirm the advantage of the COS method in terms of computational time, coupled with an increase in accuracy of at least 4 orders of magnitude, when compared to MC.

\section{Toward a Scalable and General COS Framework}\label{sec: Extensions}
This paper has demonstrated that the COS method, when combined with CP decomposition and direct training in the Fourier domain, enables efficient and accurate computation of portfolio-level risk metrics. The framework performs particularly well in moderate dimensions and for portfolios composed of linear derivatives. These findings provide a strong foundation for a deterministic, Fourier-based alternative to MC simulation.

However, practical applications often demand scalability to higher dimensions, support for exotic derivatives, and compatibility with existing simulation workflows. In this section, we outline how the COS framework can be extended to meet these demands. Each proposed direction addresses a specific limitation observed in the present work and aims to broaden the applicability of the COS method without compromising its deterministic efficiency.

\subsection{Scalability via Tensor Train Decomposition}
\label{subsec:TT-COS}
The CP decomposition used in this paper provides a simple and interpretable low-rank structure for representing the cosine coefficient tensor. It is particularly well suited to moderate-dimensional problems, where its transparent format facilitates efficient optimization via ALS and aligns naturally with the separability of cosine expansions \cite{kolda2009tensor}. These properties motivated our choice of CP in the current framework.

However, in high-dimensional settings, CP decomposition becomes increasingly fragile. Its representational power and numerical stability degrade with dimensionality, often requiring higher ranks to maintain accuracy. Our numerical experiments in Figure~\ref{fig: convergence rank} reveal that the CPD-based optimizer may converge to suboptimal solutions due to the non-convex and ill-posed nature of the problem.

To overcome these limitations, we propose replacing CP with TT decomposition in future work. This change would be implemented in the offline training component of our framework, as illustrated in Figure \ref{fig: COS framework representation}, where the CPD formulation would be substituted with the TT format. The TT format represents a high-dimensional tensor as a chain of interconnected three-dimensional cores and scales linearly with dimension \cite{oseledets2011tensor}. This structure improves numerical stability and has proven effective in option pricing applications involving 20 or more risk factors \cite{glau2020low}.

Although the TT format is more complex and less interpretable than CP, it offers the scalability and accuracy needed for large-scale portfolio evaluation. Its integration into the COS framework will allow us to extend our method to even higher-dimensional market models while retaining deterministic efficiency.


\subsection{Support for Exotic and Path-Dependent Products}
While the current implementation focuses on portfolios of linear instruments, real-world portfolios often contain exotic derivatives with path dependencies, early exercise rights, or nonlinear payoffs. These features complicate Fourier-based integration, as the pricing functions are no longer available in closed form.

To accommodate such instruments, we propose integrating pricing function approximations into the COS framework, as illustrated by the dashed arrows in Figure \ref{fig: COS framework representation}. This approach is well established in MC simulation, where surrogate models, such as Chebyshev interpolants or neural networks, are trained offline and evaluated at simulated market scenarios. In the COS setting, these surrogates can instead be evaluated at deterministic grid points. This preserves the structure of the COS integration while extending its coverage to exotic products.

A particularly appealing feature of interpolation-based surrogates is their tendency to yield separable approximations across risk factors. As discussed in Section~\ref{sec: COS for CCR}, this separability reduces the effective dimensionality of the integration problem and aligns naturally with tensor-structured methods.

This extension not only enhances the expressiveness of the COS framework but also leverages a vast body of research from MC simulation. By reusing well-established surrogate modeling techniques, the transition to COS becomes both practical and low-overhead. Future work will focus on systematically adapting these tools to deterministic grid-based integration, enabling efficient and accurate exposure evaluation for portfolios with complex derivatives.

\subsection{COS-CPD for Importance Sampling in MC}
\label{subsec:COSIS}
Although the COS method offers a deterministic alternative to MC simulation, it can also enhance MC-based methods via variance reduction. This is particularly useful in high-dimensional contexts where MC remains dominant due to its flexibility.

One promising application is within Cross-Entropy Adaptive Importance Sampling (CE-AIS) \cite{el2021improvement}, which iteratively refines an auxiliary distribution to reduce estimator variance. COS-based PFE approximations can provide fast, accurate guidance for updating the proposal distribution, especially for portfolios dominated by linear instruments.

Our numerical experiments demonstrate that the COS method is highly effective for such portfolios. In institutional settings, these instruments constitute a substantial portion of exposure. Embedding COS within CE-AIS provides a strong starting point for improving simulation efficiency.

For very high-dimensional portfolios, we propose a decomposition strategy: the portfolio is divided into lower-dimensional sub-portfolios, each evaluated with COS. Their marginal distributions are then combined using copulas to approximate the joint loss distribution. While this introduces approximation error, it suffices for guiding the importance sampling process.

This hybrid strategy allows institutions to leverage existing MC infrastructure while incorporating COS-based improvements, offering a flexible path toward more efficient risk quantification.
\section{Conclusion}
\label{sec: Conclusion}
This paper initiates a series of studies that develop a COS--tensor framework as a deterministic alternative to MC simulation for CCR quantification in large and liquid portfolios with a modest number of dominant risk factors. The key methodological shift is to reformulate exposure distribution recovery in the Fourier domain: once the characteristic function of a portfolio-level MtM value or exposure is available, the corresponding CDF (and hence tail measures such as PFE) can be recovered efficiently via the one-dimensional COS method.

To mitigate the curse of dimensionality in characteristic-function evaluation, we introduced a dimension-reduced cosine expansion framework in which the multidimensional Fourier--cosine coefficient tensor of the joint density is approximated using a low-rank tensor decomposition. Focusing on CPD for its simplicity and interpretability, we further proposed an offline training algorithm that operates directly in the Fourier domain by evaluating sampled coefficient fibers on-the-fly from the joint characteristic function. Relative to gradient-based regression in the physical domain, this Fourier-domain training yields speed and accuracy improvements of more than two orders of magnitude in the training stage.

The resulting COS--CPD framework was applied to netting-set--level MtM and exposure calculations. Numerical experiments on netting sets comprising tens of thousands of trades driven by seven risk factors show that the proposed method achieves relative errors below $0.1\%$ while requiring only a small fraction of the runtime of MC simulation. 

The main remaining computational bottleneck lies in the offline training stage, where the curse of dimensionality continues to affect scalability as the number of risk factors increases. This motivates three natural extensions. First, replacing CPD with TT decompositions and leveraging the efficient cross-approximation algorithms developed for TT formats would enable substantially higher-dimensional settings, particularly for margined netting sets. Second, broadening the practical scope of the framework requires extending it to exotic and path-dependent derivatives. This can be achieved by incorporating instrument-level acceleration techniques originally developed within MC frameworks, thereby retaining product flexibility while preserving the efficiency of the Fourier-based methodology. Third, COS--tensor approximations can be embedded within importance sampling schemes. Such a hybrid strategy extends the approach beyond liquid portfolios and facilitates counterparty-level calculations when a direct low-rank COS representation is computationally infeasible.

\printbibliography[heading=bibintoc]

\appendix
\section{Additional numerical tests}
This appendix reports COS-only results for counterparty-level PFE in the original three-dimensional setting. We include these results for completeness, since counterparty aggregation across multiple netting sets removes the CDF transformation available at the netting-set level (explained in Section~\ref{subsec:chf_nettingset_exposure}) and requires evaluating the characteristic function with the positive-part operator retained, leading to a fully three-dimensional numerical integral.

\subsection{PFE at the counterparty level}\label{subsec: PFE counterparty}
To construct a counterparty-level exposure, we reuse the portfolio from the netting-set experiments and partition it by contract type, assigning each subset to a distinct netting set. The counterparty exposure is then the sum of netting-set exposures. In contrast to the netting-set case, the aggregation prevents computing the exposure CDF via a simple transformation of the MtM CDF and therefore necessitates working directly with the exposure characteristic function; see Section~\ref{subsec: CDF for Counterparty}. The presence of the positive-part operator introduces a non-smooth payoff, which yields Gibbs-type oscillations when reconstructing distributions via truncated Fourier--cosine series.

To reduce these oscillations, we apply a spectral filter to the Fourier coefficients prior to CDF evaluation. Specifically, we use an exponential filter of order $p=4$ with decay parameter $\alpha=-\log(\epsilon_m)$, where $\epsilon_m$ denotes machine precision \cite{gottlieb1997gibbs}.

\begin{figure}[h!]
	\centering
	\begin{subfigure}[b]{0.49\textwidth}
		\centering
		\includegraphics[width=\textwidth]{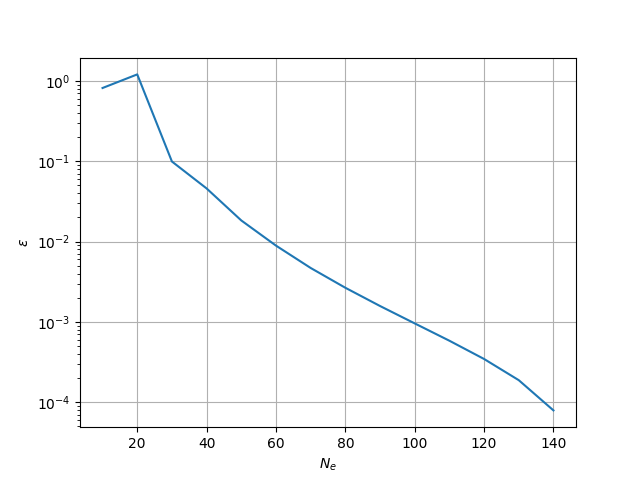}
		\caption{Relative error vs.\ number of expansion terms $N_e$.}
		\label{fig: convergence counterparty expansion}
	\end{subfigure}
	\hfill
	\begin{subfigure}[b]{0.49\textwidth}
		\centering
		\includegraphics[width=\textwidth]{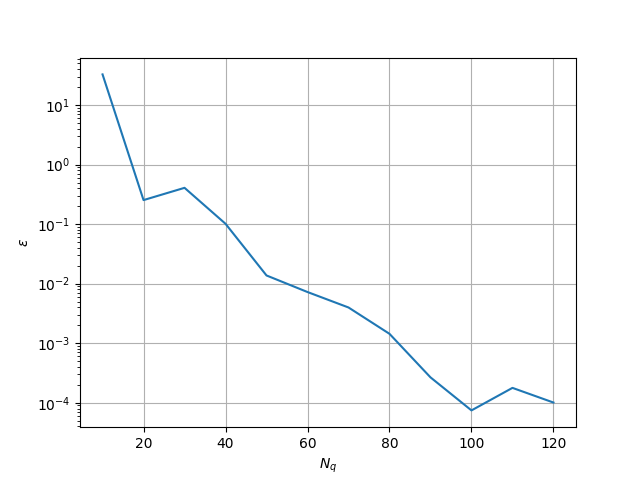}
		\caption{Relative error vs.\ number of quadrature points $N_q$.}
		\label{fig: convergence counterparty quadrature}
	\end{subfigure}
    \caption{Counterparty-level PFE in the 3D COS setting for a portfolio of $1000$ IR and FX derivatives. The plotted quantity is the relative error (in \%) with respect to the COS reference PFE value $\$11835.52$, shown on a logarithmic scale.}
	\label{fig: convergence counterparty}
\end{figure}

For the convergence study, we compute a high-accuracy COS reference using $N_e=150$ expansion terms and $N_q=130$ quadrature points with $\mathrm{TOL}=10^{-12}$, and evaluate convergence at $t=8.6$ years. Figure~\ref{fig: convergence counterparty} reports the relative error in the PFE estimate as a function of $N_e$ and $N_q$.

Compared with the netting-set experiments, convergence in $N_e$ is slower and consistent with algebraic decay after filtering. We also observe non-monotonic behaviour in the quadrature convergence with respect to $N_q$, which is likely influenced by residual oscillations in the Fourier reconstruction and their interaction with the subsequent Newton--Raphson root-finding step.

Table~\ref{table:comptime_counterparty_combined} reports runtime and accuracy for COS and MC using the same exposure dates and COS settings as in the netting-set case to enable direct comparison.

\begin{table}[h!]
\caption{Accuracy and computational time required to calculate the PFE of 3-dimensional counterparty-level portfolios with $1000$ and $10,000$ derivatives. Relative errors (in $\%$) are averaged over two exposure dates.}
\label{table:comptime_counterparty_combined}
\centering
\begin{tabular}{@{} l r S[table-format=1.2e-1, round-mode=places, round-precision=2] @{}}
\toprule
Method & {CPU Time (s)} & {Error (\%)} \\
\midrule
\multicolumn{3}{l}{\emph{Portfolio with 1,000 derivatives}} \\
		MC $(0.5 \cdot 10^5)$ & 1.6            & 0.251              \\
		MC $(10^5)$           & 3.1           & 0.172             \\
		MC $(5 \cdot 10^5)$   & 16.9            & 0.094            \\
		COS                 & 0.6             & 0.071             \\
\midrule
\multicolumn{3}{l}{\emph{Portfolio with 10,000 derivatives}} \\
		MC $(0.5 \cdot 10^5)$ & 13.3             & 0.709            \\
		MC $(10^5)$           & 26.5            & 0.449            \\
		MC $(5 \cdot 10^5)$   & 147.5           & 0.165           \\
		COS                 & 1.2             & 0.061              \\
\bottomrule
\end{tabular}
\end{table}

The results in Table~\ref{table:comptime_counterparty_combined} show that COS remains computationally competitive at the counterparty level under aggregation across netting sets. The exponential filter adds negligible overhead relative to the overall COS runtime in these experiments. Relative to the netting–set case, the COS error increases by approximately two orders of magnitude.
Nonetheless, COS remains superior to MC in both runtime and accuracy across all tested configurations.

\end{document}